\documentclass[twocolumn]{aastex62}
\usepackage{amsmath}
\usepackage{wasysym}
\usepackage{savesym}
\savesymbol{tablenum}
\usepackage{siunitx}
\restoresymbol{SIX}{tablenum}
\usepackage{booktabs}

\DeclareSIUnit\Mearth{M_\oplus}
\DeclareSIUnit\Rearth{R_\oplus}
\DeclareSIUnit\Mmoon{M_\textrm{\leftmoon}}
\DeclareSIUnit\JEM{J_{EM}}

\graphicspath{{./}{figures/}}

\received{04.09.2024}
\revised{04.11.2024}
\accepted{08.11.2024}
\submitjournal{ApJ}

\shorttitle{Survey of Moon-Forming Impacts}
\shortauthors{Meier et al.}

\begin{document}

\title{A Systematic Survey of Moon-Forming Giant Impacts. II. Rotating bodies}

\correspondingauthor{Thomas Meier}
\email{thomas.meier5@uzh.ch}

\author[0000-0001-9682-8563]{Thomas Meier}
\affiliation{Department of Astrophysics, University of Zurich, Winterthurerstrasse 190, CH-8057 Zurich, Switzerland}

\author[0000-0002-4535-3956]{Christian Reinhardt}
\affiliation{Department of Astrophysics, University of Zurich, Winterthurerstrasse 190, CH-8057 Zurich, Switzerland}
\affiliation{Physics Institute, Space Research and Planetary Sciences, University of Bern, Sidlerstrasse 5, CH-3012 Bern, Switzerland}

\author[0000-0003-1938-7877]{Miles Timpe}
\affiliation{Department of Astrophysics, University of Zurich, Winterthurerstrasse 190, CH-8057 Zurich, Switzerland}

\author[0000-0001-7565-8622]{Joachim Stadel}
\affiliation{Department of Astrophysics, University of Zurich, Winterthurerstrasse 190, CH-8057 Zurich, Switzerland}

\author[0000-0001-5996-171X]{Ben Moore}
\affiliation{Department of Astrophysics, University of Zurich, Winterthurerstrasse 190, CH-8057 Zurich, Switzerland}



\begin{abstract}
In the leading theory of lunar formation, known as the giant impact hypothesis, a collision between two planet-size objects resulted in a young Earth surrounded by a circumplanetary debris disk from which the Moon later accreted. The range of giant impacts that could conceivably explain the Earth-Moon system is limited by the set of known physical and geochemical constraints. However, while several distinct Moon-forming impact scenarios have been proposed---from small, high-velocity impactors to low-velocity mergers between equal-mass objects---none of these scenarios have been successful at explaining the full set of known constraints, especially without invoking one or more controversial post-impact processes. Allowing for pre-impact rotation of the colliding bodies has been suggested as an avenue which may produce more promising collision outcomes. However, to date, only limited studies of pre-impact rotation have been conducted. Therefore, in the second paper of this series, we focus on pairwise impacts between rotating bodies. Using non-rotating collisions as a baseline, we systematically study the effects of rotation on collision outcomes. We consider nine distinct rotation configurations and a range of rotation rates up to the rotational stability limit. Notably, we identify a population of collisions that can produce low post-impact angular momentum budgets and massive, iron-poor protolunar disks. Furthermore, even when pre-impact rotation is included, we demonstrate that the canonical Moon-forming impact can only generate sufficiently massive protolunar disks in the presence of excessive post-impact angular momentum budgets; this casts doubt on the canonical impact scenario.
\end{abstract}



\section{Introduction} \label{sec:intro}
The prevailing theory on the formation of Earth's Moon is the Giant Impact (GI) hypothesis. It proposes that a collision between the young Earth and a planetary-sized body ejected material into a circumplanetary disk from which the Moon then formed \citep{hartmannSatelliteSizedPlanetesimalsLunar1975,cameronOriginMoon1976}. In the leading version of this hypothesis, called the "canonical" Moon-forming impact, the impactor is roughly Mars-sized and the collision is oblique and occurs at a low impact velocity, $v_{imp}\simeq v_{esc}$, where $v_{esc}$ is the mutual escape velocity of the two bodies.

A successful Moon-forming collision must satisfy a number of known constraints. As these constraints are already discussed at length in \citet{timpeSystematicSurveyMoonforming2023} (hereafter Paper I), here we only reiterate those constraints which are directly relevant to our simulations. First, such a collision has to eject at least one lunar mass of material into orbit to allow the formation of the Moon. The proto-lunar disk also has to be strongly depleted in iron to explain the small iron core of the Moon which is $\leq \SI{1.5}{\percent}$ by mass \citep{williamsLunarInteriorProperties2014}. Then the total angular momentum of the Earth-Moon system has to be consistent with the observed value (\SI{}{\JEM}). Finally, the Earth and the Moon have an indistinguishable isotopic composition in several elements including $^{18}O/^{17}O$ \citep{wiechertOxygenIsotopesMoonForming2001}, $^{50}Ti/^{47}Ti$ \citep{zhangProtoEarthSignificantSource2012}, and $^{182}W/^{184}W$ \citep{touboulLateFormationProlonged2007}. This in turn implies that either the proto-Earth and the impactor initially had a very similar isotopic composition (and therefore formed at a similar heliocentric distance) or that the material of the two colliding bodies was very well mixed during the collision.

Early simulations \citep{canupOriginMoonGiant2001, canupSimulationsLateLunarforming2004} suggest that the canonical scenario could eject approximately one lunar mass of material into orbit while simultaneously reproducing the observed angular momentum (AM) of the Earth-Moon system and the low iron content of the Moon. However, most of the material that forms the protolunar disk is derived from the impactor and it therefore exhibits poor mixing. As a consequence, new scenarios were proposed to reconcile the GI hypothesis with isotopic constraints. One of these scenarios, first proposed by \citet{canupFormingMoonEarthlike2012}, is a merger of near-equal mass bodies. This scenario can produce near-perfect mixing due to the symmetry of the impact. Another scenario that can result in a well-mixed protolunar disk is a high-velocity impact by a small impactor onto a rapidly spinning proto-Earth \citep{cukMakingMoonFastSpinning2012}. This latter scenario can recover the isotopic similarity by ejecting material primarily from the proto-Earth into orbit. However, both of these scenarios result in excess angular momentum of \SIrange{1}{2}{\JEM}. It is still being debated by how much the two proposed post-impact processes can reduce the angular momentum of the Earth-Moon system. Estimates for the solar evection resonance range from a few percent \citep{tianCoupledOrbitalthermalEvolution2017} to several \SI{}{\JEM} \citep{cukMakingMoonFastSpinning2012} depending on the underlying tidal model, while the Laplace plane transition can, depending on the Earth's initial obliquity, reduce the initial angular momentum by a factor of two to three \citep{cukTidalEvolutionMoon2016, cukTidalEvolutionEarthMoon2021}.

During the planet formation process, terrestrial planets are expected to rotate rapidly due to accretion of small bodies and giant impacts (e.g., \citealt{agnorCharacterConsequencesLarge1999,kokuboFORMATIONTERRESTRIALPLANETS2010,quintanaFREQUENCYGIANTIMPACTS2016}). Accounting for different pre-impact spins of the colliding bodies when investigating the Moon-forming GI will broaden the parameter space and expand the range of collision outcomes. However, most prior work on the GI hypothesis has been limited to initially non-rotating bodies. So far, only a few studies, e.g., \citet{canupLunarformingCollisionsPreimpact2008}, \citet{cukMakingMoonFastSpinning2012}, \citet{ruiz-bonillaEffectPreimpactSpin2021}, and \citet{kegerreisImmediateOriginMoon2022}, have investigated impacts between rotating bodies but those were limited to (relatively) narrow regions of the giant impact parameter space, such as the canonical impact or high-velocity impacts on a rapidly spinning proto-Earth. In this paper, we investigate collisions with pre-impact rotation of both the target and the impactor.

As it stands, no giant impact scenario has been shown to simultaneously reproduce all known constraints of the Earth-Moon system without requiring very specific assumptions regarding post-impact processes or the initial composition of the colliding bodies. Moreover, prior work has largely focused on a limited range of impact parameters in order to explain specific observational constraints. A systematic investigation of the parameter space of potential Moon-forming impacts has so far not been performed.

Therefore, in this work, we present a systematic survey of Moon-forming giant impacts. The aim of this study is to provide the community with a comprehensive survey of the parameter space and a systematic analysis of the collision outcomes. The simulations in this study assume a single giant impact event and the subsequent post-impact analysis determines whether any such event can simultaneously explain the observed physical, compositional, and geochemical constraints of the Earth-Moon system. 

We have chosen to split the results into two papers in order to keep the results tractable. Paper I focused on the subset of collisions \emph{without} pre-impact rotation and provides a baseline against which the effects of pre-impact rotation can be compared. In Paper I, we found that, in order to obtain a sufficiently massive protolunar disk (e.g., $M_d \geq \SI{}{\Mmoon}$) without pre-impact rotation, an initial angular momentum budget of at least two times the current value of the Earth-Moon system ($J_0 \gtrsim \SI{2}{\JEM}$) is required. Therefore, without pre-impact rotation, a post-impact process capable of removing at least one \SI{}{\JEM} is needed to reconcile such collisions with the observational constraint. This also clearly refutes the canonical scenario, as it does not produce a disk massive enough to form the Moon. Furthermore, Paper I also demonstrated that good mixing between proto-Earth and impactor material can only be consistently realized in low-velocity collisions between near-equal mass bodies, as proposed in \citet{canupFormingMoonEarthlike2012}. This type of equal-mass merger together with a post-impact process that is able to remove at least \SI{1}{\JEM} would thus be able to explain the formation of the Moon.

In the present paper (hereafter Paper II), we broaden the parameter space and consider collisions \emph{with} pre-impact rotation of the proto-Earth and impactor for a wide range of rotational configurations. The expanded parameter space adds 7152 collisions to the data set presented in Paper I. Adding pre-impact rotation to the colliding bodies introduces six new degrees of freedom (i.e., two angles for the orientation of the spin axis and a rotation rate for each body). We show that these new impacts fill some of the regions of the post-impact parameter space that were left empty by the non-rotating collisions studied in Paper I, e.g., by producing sufficient disk masses at lower post-impact angular momentum budgets. But considering these additional impacts also produces degeneracy in the disk parameters, meaning that vastly different initial conditions can produce very similar post-impact disks.

This paper is structured as follows: in Section~\ref{sec:methods}, we will first reiterate the methods used to perform the impact simulations and the subsequent analysis, while emphasizing the differences to Paper I. In Section~\ref{sec:results_and_discussion}, the results of all impact simulations, including those already studied in Paper I, are presented and discussed. Finally, in Section~\ref{sec:summary_and_conclusions} we provide a summary of our findings and present their implications for the giant impact hypothesis. In Appendix \ref{sec:appendix_parameter_space}, the pre-impact parameter space is described in detail. In Appendix \ref{sec:correlation_full_data_set}, we investigate the correlations between selected pre- and post-impact parameters. In Appendix \ref{sec:appendix_immediate}, we explore the concept of immediately formed satellites.

\section{Methods} \label{sec:methods}
The methods used to perform the impact simulations and their analysis follow the procedures described in Section 3 of Paper I, thus in this section we highlight the differences to Paper I. We use the Smoothed Particle Hydrodynamics (SPH) code \texttt{Gasoline} \citep{wadsleyGasolineFlexibleParallel2004} with modifications for giant impact simulations \citep{reinhardtNumericalAspectsGiant2017, reinhardtBifurcationHistoryUranus2020, meierEOSResolutionConspiracy2021}. After the impact, the post-impact state of the system is analyzed with \texttt{SKID} \citep{n-bodyshopSKIDFindingGravitationally2011} to identify gravitationally bound fragments and classify the outcome of the collision. For collisions that result in a merger, the planet, circumplanetary disk, and ejecta are differentiated using the novel disk finder presented in Paper I. A detailed description of the disk finding algorithm can be found in Appendix C of Paper I and the code implementation is freely available on GitHub at \citet{timpeDisk_finder_timpe2023}.

\subsection{Rotating pre-impact models} \label{sec:method-rotation}
In the present paper, we consider pre-impact rotation of the target (i.e., the proto-Earth) and impactor. This introduces additional steps when generating the initial planet models. In order to achieve rotation in the targets and impactors, we follow the approach introduced in \citet{timpeMachineLearningApplied2020}. First, following the same procedure explained in Section 3.2 of Paper I, we create a non-rotating model with the desired mass, Earth-like composition (iron core \SI{33}{\percent} and rocky mantle \SI{67}{\percent} by mass) and surface temperature ($T_{s} = \SI{1000}{\kelvin}$). We then evolve the particle representation of the model in a co-rotating coordinate frame and gradually increase the centrifugal force until the desired (uniform) angular velocity is achieved.

The pre-impact rotation rate is parameterized using the critical angular velocity, which is the angular velocity at which the body is expected to become rotationally unstable,

\begin{equation} \label{eq:omega_crit}
    \Omega_{crit} = \sqrt{\pi G \bar{\rho} h_{crit}} \,,
\end{equation}

\noindent where $G$ is Newton's gravitational constant, $\bar{\rho}$ is the bulk density of the non-rotating model, and $h_{crit} = 0.44931$ is derived from MacLaurin's formula \citep{chandrasekharEllipsoidalFiguresEquilibrium1969,ansorgUniformlyRotatingAxisymmetric2003}. The initial angular velocity of each model is therefore parameterized as,

\begin{equation} \label{eq:scaled-rotation-rate}
   \Omega=f_{\Omega}\Omega_{crit} \,,
\end{equation}

\noindent where $f_{\Omega}$ is a scalar value called the angular velocity factor.

The model is then transferred from the co-rotating frame into the stationary frame by adding the velocity components corresponding to the solid body rotation to each particle and then rotating the body to the desired rotation orientation.

\subsection{Initial conditions}\label{sec:initial_conditions}
The initial conditions for each collision are generated as described in Paper I, with the addition of six free parameters, i.e. the rotation rate and the orientation of the rotation axis of the two bodies. As in Paper I, the initial total mass ($M_{tot}$) in every collision is \SI{1.05}{\Mearth}. The masses of the target and impactor are determined by $M_{tot}$ and $\gamma$,

\begin{equation}
M_{targ} = M_{tot} \left( \frac{1}{\gamma + 1} \right)\,,\label{eq:target-mass}
\end{equation}
\begin{equation}
M_{imp} = M_{tot} \left( \frac{\gamma}{\gamma + 1} \right)\,.\label{eq:impactor-mass}
\end{equation}

\noindent The total number of particles in each collision is set to \SI{100000}{} and the particles are distributed amongst the target and projectile in proportion to their mass.

Sampling the six additional free parameters introduced by pre-impact rotation (i.e., two angles for the orientation of the spin axis and a rotation rate for each body) with any reasonable resolution would lead to an infeasible number of simulations. Thus, in order to enable a tractable systematic study, we reduce these six parameter to two by constraining the mutual orientation of the colliding bodies' angular momentum vectors to three distinct configurations. The rotational state of each body is specified relative to the angular momentum vector of the collision's orbit, which always points in the positive $z$-direction. The possible states are: the body is not rotating (N; for ``non-rotating''), the body's angular momentum vector is oriented parallel (U; for ``up'') or anti-parallel (D; for ``down'') to the collision's orbital angular momentum vector. Thus, all angular momentum vectors ($\vec{J}_0$, $\vec{J}_{orb}$, $\vec{J}_{targ}$, $\vec{J}_{imp}$) point in either the positive or negative $z$-direction and can be reduced to scalars ($J_0$, $J_{orb}$, $J_{targ}$, $J_{imp}$).

From the three possible spin orientations of the bodies (N/U/D), nine different configurations are possible (because the angular velocity factor $f_\Omega$ is always the same for both bodies except when one is zero because it is non-rotating): the target and impactor are both non-rotating (NN), the target and impactor are both rotating with their angular momentum vectors pointing upwards (UU), the target is rotating with its angular momentum vector pointing upwards and the impactor is non-rotating (UN), the target is non-rotating and impactor is rotating with its angular momentum vector pointing upwards (NU), and so on. Note that the first letter in this notation always indicates the rotational state of the target and the second letter the rotational state of the impactor.

Of the 7649 simulations performed for this study, 497 are non-rotating (NN) collisions. These collisions were extensively discussed in Paper I. For the remaining rotational configurations, simulations are performed with the pre-impact bodies rotating at different rotation rates set by Equation~\eqref{eq:scaled-rotation-rate}.

In contrast to Paper I, wherein $J_0$ is simply the orbital angular momentum (of the impactor's orbit around the target), $J_0$ in this paper is the sum of the collision's orbital angular momentum (motion of the impactor relative to the target) and the rotational angular momentum of the target and impactor (spin). Thus, for a given rotational configuration and value of $J_0$, the orbital angular momentum ($J_{orb}$) can be obtained from

\begin{equation}
J_{0} = J_{orb} + J_{targ} + J_{imp} \,, \label{eq:J-budget}
\end{equation}

\noindent where the angular momenta are measured in the collision's center of mass frame. The pre-impact angular momenta $J_0$, $J_{targ}$, and $J_{imp}$ are therefore independent parameters in the initial conditions while $J_{orb}$ depends on these other parameters. With this, we can calculate the asymptotic impact parameter ($b_{\infty}$) similar to Paper I (note the change from $J_0$ to $J_{orb}$):

\begin{equation}
b_{\infty} = \frac{J_{orb}}{M_{tot} v_{\infty}} \frac{\left( \gamma + 1 \right)^{2}}{\gamma} \,.\label{eq:com-binf}
\end{equation}

From this point onward, the procedure is exactly the same as in Paper I, with the only exception being that for the radii of the bodies we use the equatorial radii (i.e. the largest distance between the center of the body and any particle). This means that the definition of $R_{crit}$ used in Paper I is replaced with $R_{crit} = R_{targ,eq} + R_{imp,eq}$, using the equatorial radii of rotating bodies and normal radii for non-rotating bodies.

The total length of each simulation ($\tau$) is the sum of the pre-impact phase ($\tau_{pre}$)---which depends on the initial pre-impact state (e.g., $v_{\infty}$) and is determined analytically---and the post-impact phase ($\tau_{post}$) which is fixed. In this study, the post-impact phase is equivalent to $\tau_{post} = 7$ days. In some cases, graze-and-merge encounters have not resolved within this time limit. In these cases, which are rare, the simulation is continued in blocks of 7 days, until the encounter has either resolved or a maximum in-simulation time of 42 days is reached. Those simulations that do not resolve within 42 days are excluded from the analysis.

In summary, the free parameters in this study are the rotational configuration, total initial angular momentum $J_0$, asymptotic relative velocity $v_{\infty}$, impactor-to-target mass ratio $\gamma$ and angular velocity factor $f_\Omega$. The experimental design of the resulting parameter space is described in detail in Appendix \ref{sec:appendix_parameter_space}.

\section{Results \& Discussion} \label{sec:results_and_discussion}
\begin{table}
\centering
\begin{tabular}{lrrrrr}
    \toprule
    Config. & Runs & \multicolumn{2}{c}{Merger with $M_b$} & H\&R & G\&M\\
    &&$\geq\SI{1.0}{\Mearth}$&$<\SI{1.0}{\Mearth}$&&\\
    \midrule
    DD & 638 & 546 & 4 & 88 & 0\\
    DN & 1681 & 1321 & 62 & 297 & 1\\
    DU & 608 & 461 & 31 & 114 & 2\\
    ND & 594 & 541 & 0 & 53 & 0\\
    NN & 497 & 335 & 20 & 142 & 0\\
    NU & 622 & 550 & 11 & 60 & 1\\
    UD & 667 & 637 & 2 & 26 & 2\\
    UN & 1724 & 1267 & 96 & 361 & 0\\
    UU & 618 & 589 & 0 & 29 & 0\\
    \midrule
    Total & 7649 & 6247 & 226 & 1170 & 6\\
    \bottomrule
\end{tabular}
\caption{Overview of the simulation set. A grand total of 7649 collisions were simulated for this study (including the simulations from Paper I). Of these, 6247 are considered in the analysis. Of the remaining 1402 simulations, 226 result in an insufficient bound mass $M_b<\SI{1.0}{\Mearth}$, 1170 are hit-and-run collisions (H\&R) which are excluded from the analysis (see Paper I for details) and 6 are unresolved graze-and-merge encounters (G\&M).}
\label{tab:Simulation_Overview}
\end{table}

\begin{figure*}[ht!]
\centering
\plotone{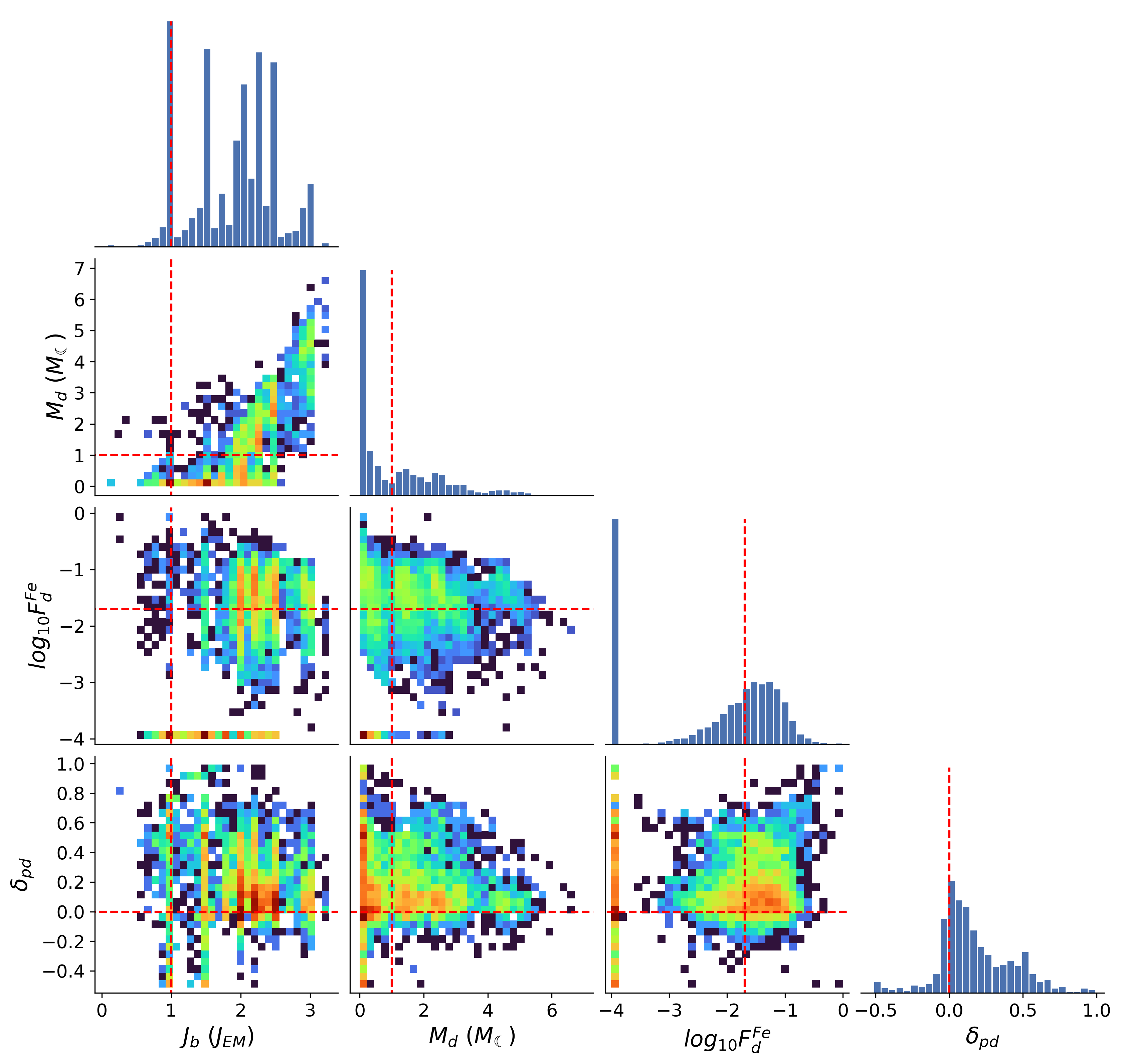}
\caption{Correlation plot between the main post-impact variables. Shown are the bound angular momentum $J_b$, disk mass $M_d$, disk iron mass fraction $F^{Fe}_{d}$ and compositional difference $\delta_{pd}$ for all 6247 simulations considered in the analysis (for details see the caption of Table~\ref{tab:Simulation_Overview}). The panels on the diagonal show univariate histograms with linear counts, while the lower-left triangle shows bivariate histograms with logarithmic counts. The dashed red lines mark the constraints discussed in the introduction ($J_b=\SI{1}{\JEM}$, $M_d=\SI{1}{\Mmoon}$, $F_d^{Fe}=\SI{2}{\percent}$ and $\delta_{pd}=\SI{0}{}$). Impacts that do not produce any disk ($M_d = \SI{0}{\Mmoon}$) are excluded from panels involving $F_d^{Fe}$ or $\delta_{pd}$ because these values do not have meaning if there is no disk. As the disk iron mass fraction $F_d^{Fe}$ is shown on a logarithmic axis, the 1478 simulations that result in exactly zero disk iron mass fraction (even though the disk mass can be significant) would be invisible, so they are moved to a value of \SI{1e-4}{} throughout all plots in this paper (the lowest non-zero value is \SI{1.7e-4}{} so they can be clearly distinguished). For the impact conditions considered in this study, collision outcomes are very diverse. The pronounced peaks in the distribution of the bound angular momentum $J_b$ are caused by the sampling of the initial angular momentum $J_0$ in the initial conditions (see Appendix~\ref{sec:appendix_parameter_space}).}
\label{fig:Seaborn_Triangle_plot_only_output_variables}
\end{figure*}

In Table \ref{tab:Simulation_Overview}, we provide an overview of the collision outcomes, including the results of the non-rotating simulations presented in Paper I. Out of the grand total of 7649 simulations that were performed, 6247 are considered in our analysis as potential Moon-forming impacts, while 1402 are rejected for various reasons. Of the 1402 simulations that are rejected, 226 are rejected due to the post-impact bound mass being too small ($M_b<\SI{1}{\Mearth}$), 1170 are hit-and-run collisions which are excluded from the analysis (see Paper I for a detailed discussion), and a further six collisions are still classified as unresolved graze-and-merge encounters (i.e., the merger has yet to occur by the maximum simulation time of \SI{42}{\day} after the initial impact).

In our analysis, we focus on four post-impact properties which are either directly related to known constraints ($J_b$) or are necessary proxies to such constraints ($M_d$, $F^{Fe}_d$, $\delta_{pd}$). $J_b$ is the angular momentum budget of the bound material remaining after the impact and should match the currently observed angular momentum budget of the Earth-Moon system (\SI{}{\JEM}) or be within a range set by post-impact angular momentum removal processes. $M_d$ is the mass of the post-impact circumplanetary disk (i.e., the protolunar disk) and must be at least one lunar mass ($\SI{}{\Mmoon} = \SI{0.0123}{\Mearth}$) to provide enough material for lunar accretion; however, previous studies suggest that \SIrange{2}{4}{\Mmoon} is required under realistic accretion efficiencies. $F^{Fe}_{d}$ is the iron mass fraction of the protolunar disk, a property which will affect the iron mass fraction of the Moon which subsequently forms out of the disk. As the iron mass fraction of the Moon is constrained to less than \SI{2}{\percent}, $F^{Fe}_{d}$ should not exceed 0.02 unless differential accretion rates are invoked. $\delta_{pd}$ is the compositional difference between the post-impact planet and disk (see Equation \ref{eq:delta-mix}), where the impactor mass fraction is used as a proxy for isotopic composition.

To be considered a successful Moon-forming impact, a simulation must reproduce all four of these constraints simultaneously; if post-impact angular momentum removal (e.g., a Solar evection resonance) or compositional mixing processes (e.g., a synestia, a rapidly spinning structure where the planet is in contact with the circumplanetary disk \citep{lockOriginMoonTerrestrial2018}) are invoked, then certain constraints can be relaxed but the resulting parameter ranges must still be satisfied simultaneously. As post-impact compositional equilibriation can only occur when a synestia is present, the proximity of the post-impact planet to the hot-spin stability limit ($R_p/R_{HSSL}$) must also be considered when invoking such a process. Therefore, the correlation between these post-impact properties is crucial to identifying Moon-forming impacts and we dedicate a considerable part of the subsequent analysis to these correlations.

In Figure \ref{fig:Seaborn_Triangle_plot_only_output_variables}, the four main post-impact properties are shown for all 6247 simulations considered in this analysis. The distribution of each property can be discerned, as well as several correlations between properties. Several trends are worth remarking on. First, for the range of pre-impact conditions considered in this study (see Appendix \ref{sec:appendix_parameter_space}), the collision outcomes are very diverse. Indeed, $J_b$ ranges from \SI{0.08}-\SI{3.27} {\JEM}, with pronounced peaks at the values of $J_0$ that are more frequently sampled by the initial conditions (i.e., $1.0$, $1.5$, $2.0$, $2.25$, $2.5$ and $\SI{3.0}{\JEM}$). $M_d$ ranges from \SIrange{0}{6.71}{\Mmoon} and a significant number of simulations ($N=1166$) produce disks with masses in the favorable range of \SI{2}{\Mmoon} and \SI{4}{\Mmoon}. The impacts that do not produce any disk ($N=1022$; $M_d=\SI{0}{\Mmoon}$) result in undefined values of $F_d^{Fe}$ and $\delta_{pd}$, and are excluded from panels involving either value. $F^{Fe}_d$ spans the full range between pure rock ($F_d^{Fe}=0$) and pure iron ($F_d^{Fe}=1$), with a peak in the distribution at $F_d^{Fe}\simeq0.03$; a significant number of cases ($N=1478$) produce disks with no iron ($F^{Fe}_{d}=0$). $\delta_{pd}$ ranges from $\delta_{pd}=-0.51$ (where the disk has a lower impactor mass fraction than the planet) to $\delta_{pd}=1.00$ (where the disk is comprised entirely of impactor material) with a pronounced peak at zero (perfect mixing) and at $\delta_{pd}=0.5$ (where the disk has a higher impactor fraction than the planet).

\subsection{Relation between pre- and post-impact angular momentum budgets}\label{sec:Angular_Momentum}
\begin{figure}[ht!]
\centering
\plotone{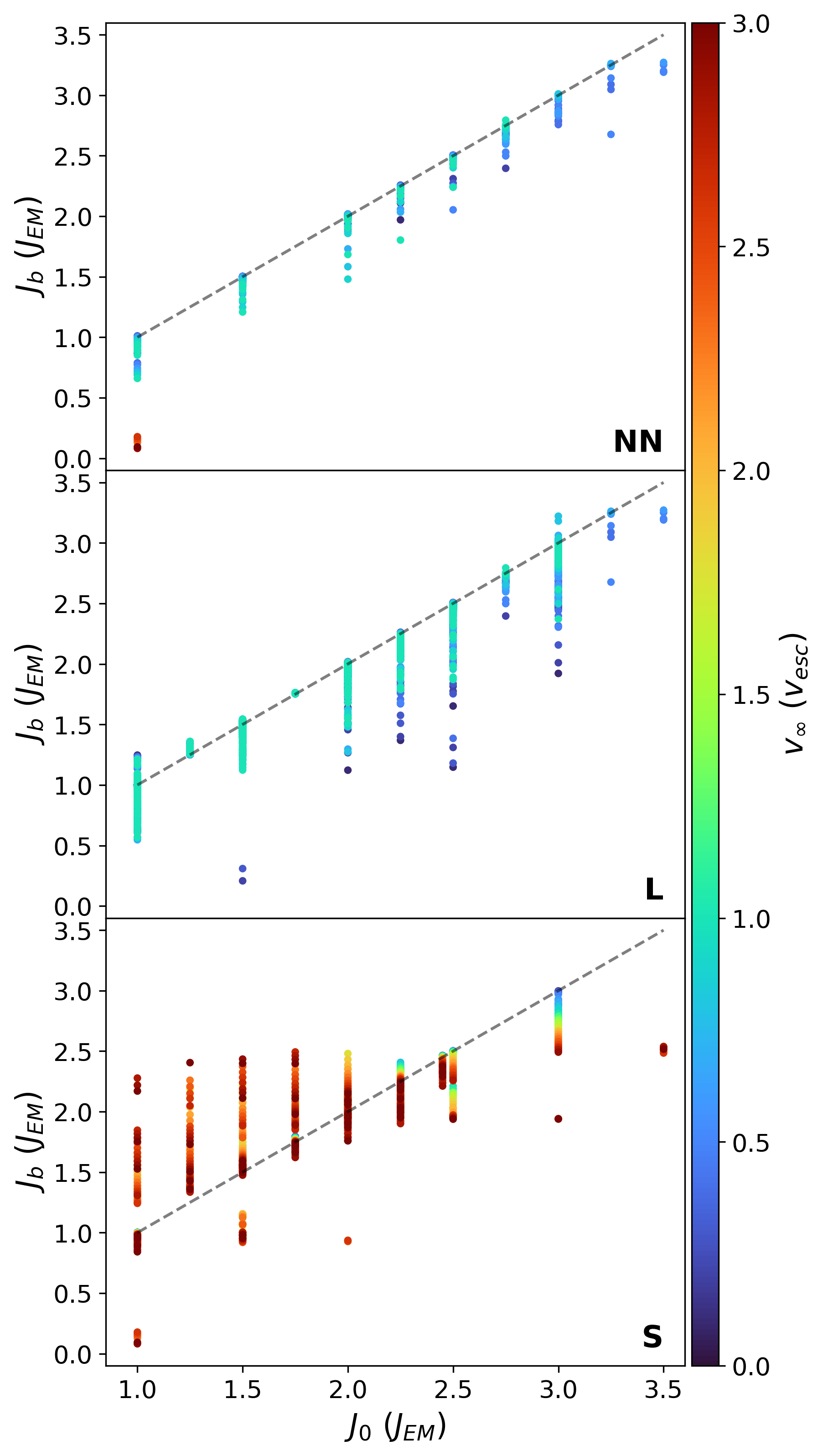}
\caption{Relation between the pre-impact ($J_0$) and post-impact bound angular momentum ($J_b$). $J_0$ is measured in the center-of-mass frame of the colliding bodies while $J_b$ is measured in the frame of the planet after the impact. The top panel shows the result of the non-rotating collisions (NN) presented in Paper I. For non-rotating bodies, there is a clear relation ($r=0.98$, see also discussion of correlations in Appendix \ref{sec:correlation_full_data_set}) between the initial and the bound angular momentum. An exception are the high velocity impacts (red dots) but those collisions do not produce massive disks and are excluded from the analysis in Paper I. The center panel contains all cases with $\gamma \geq 0.1$ (L) and shows a larger spread, in some cases the difference between $J_0$ and $J_b$ is more than \SI{75}{\percent}. It also shows some cases that deviate in the positive direction, having more bound than initial angular momentum after the collision. The bottom panel contains the low-$\gamma$ (S) cases of the full data set. This set now exhibits large deviations between $J_0$ and $J_b$ in both the positive and negative direction, where $J_b$ can be more than double or less than half the value of $J_0$. Contrary to the NN cases in Paper I, when considering the full data set $J_0$ is not a reliable predictor of the post-impact bound angular momentum anymore. We thus use $J_b$ in the analysis of the collision results in Paper II.}
\label{fig:J_comparison}
\end{figure}

We observe a strong correlation ($r=0.97$) between the pre-impact angular momentum budget ($J_0$; measured in the center-of-mass frame of the colliding bodies) and the post-impact angular momentum budget of the bound material ($J_b$; measured in the frame of the post-impact planet) as shown in Figure~\ref{fig:pearson_corr} in Appendix~\ref{sec:correlation_full_data_set}. In Paper I, we assumed that this correlation holds because little mass is lost in the collision and the associated ejecta does not carry away a significant amount of angular momentum. This is confirmed by the top panel of Figure~\ref{fig:J_comparison}, which shows the relationship between $J_0$ and $J_b$ for all non-rotating collisions. With the exception of the low-$\gamma$ simulations (i.e., small impactors) that do not produce massive disks for non-rotating collisions, all results scatter around the $J_b=J_0$ line with differences between $J_0$ and $J_b$ of less than \SI{20}{\percent}. Thus, $J_0$ is an effective proxy for $J_b$ in non-rotating collisions and, in Paper I, we used $J_0$ in the analysis of the collision outcomes.

When pre-impact rotation is introduced (center and bottom panels of Figure~\ref{fig:J_comparison}), $J_b$ and $J_0$ remain strongly correlated ($r=0.97$). Indeed, in the majority of cases ($N=5896$), the difference between $J_0$ and $J_b$ is less than \SI{20}{\percent} of $J_0$. However, in contrast to the non-rotating dataset, the rotating dataset is host to a large number of outliers, with deviations up to \SI{128.6}{\percent} of $J_0$. Thus, for the purposes of our analysis, $J_0$ is no longer a suitable proxy for $J_b$ despite its strong correlation. We directly use $J_b$ in the analysis that follows.

The bottom panel of Figure~\ref{fig:J_comparison}, which shows low-$\gamma$ collisions (i.e., small impactors), suggests that there are additional regions of the pre-impact parameter space that may be of interest to the Moon-formation community. Notably, significantly larger values of $J_b$ can be achieved for a given value of $J_0$, implying that one could investigate the region $J_0 < \SI{}{\JEM}$ to obtain results with $J_b \simeq \SI{}{\JEM}$. However, our dataset shows that the cases populating the region with $J_0 = \SI{}{\JEM}$ and $\SI{}{\JEM} < J_b < \SI{2.0}{\JEM}$ do not produce significantly massive disks ($M_{d}<\SI{0.1}{\Mmoon}$). Thus, this additional part of the pre-impact parameter space is not considered in this study but may still be explored in future work.

\subsection{Post-impact parameters}
\begin{figure*}[ht!]
\centering
\plotone{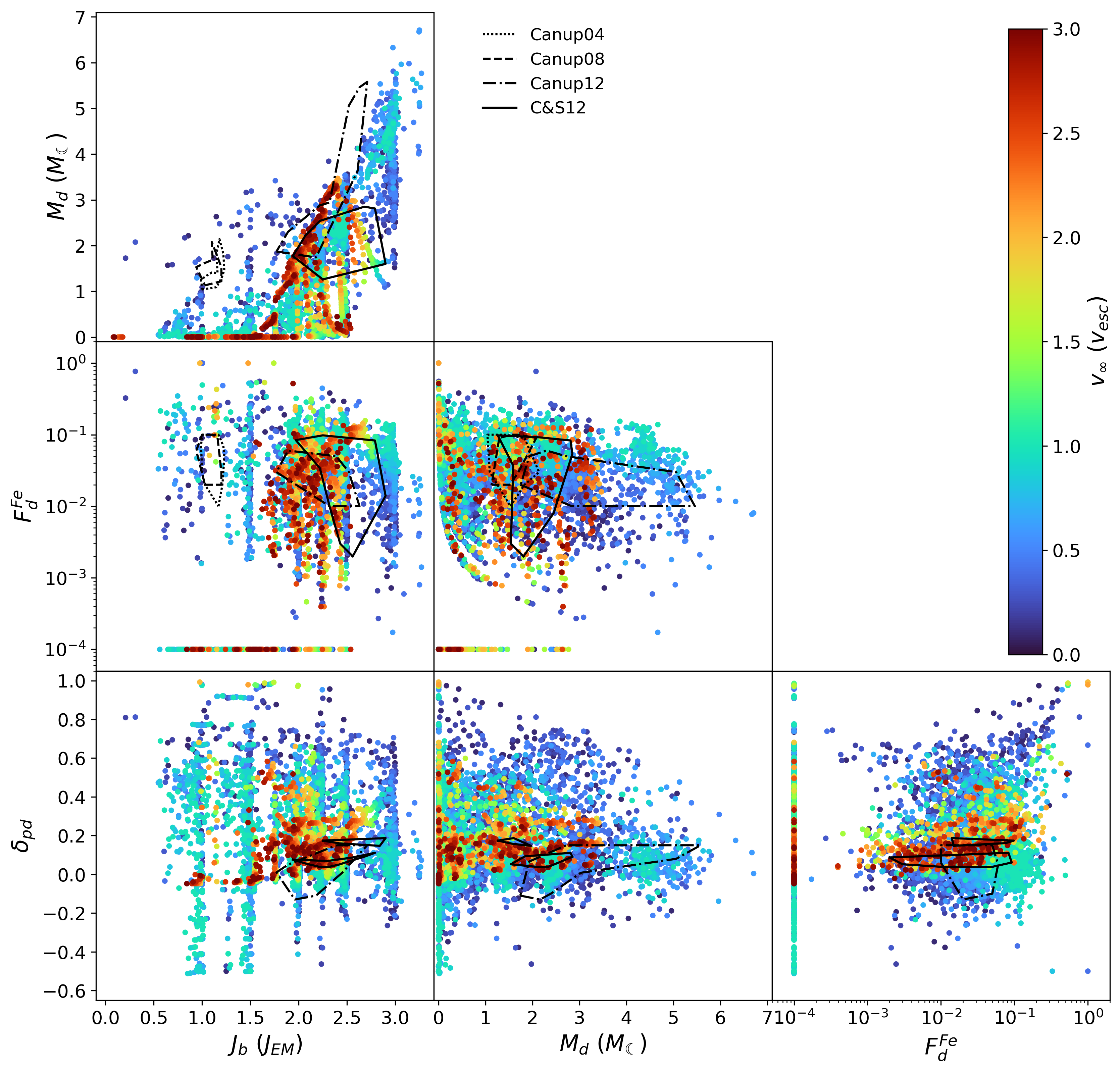}
\caption{Post-impact variables of all 6247 analyzed simulations. The figure contains six panels, each comparing two of the post-impact values with each other, arranged like Figure~\ref{fig:Seaborn_Triangle_plot_only_output_variables}. The top panel compares the disk mass $M_d$ with the bound angular momentum $J_b$. The middle row compares the disk iron mass fraction $F_d^{Fe}$ with the bound angular momentum $J_b$ and the disk mass $M_d$. The bottom row compares the mixing parameter $\delta_{pd}$ with the bound angular momentum $J_b$, the disk mass $M_d$ and the disk iron mass fraction $F_d^{Fe}$. Disk iron mass fractions of exactly zero are moved to \SI{1e-4}{} to be visible. For details see caption of Figure \ref{fig:Seaborn_Triangle_plot_only_output_variables}. The black lines mark the regions investigated in \citet{canupSimulationsLateLunarforming2004, canupLunarformingCollisionsPreimpact2008, canupFormingMoonEarthlike2012} and \citet{cukMakingMoonFastSpinning2012}.}
\label{fig:all_other_paper_data_lines}
\end{figure*}

Figure~\ref{fig:all_other_paper_data_lines} shows the results of all 6247 simulations in six panels, with each panel showing a pair of the four main post-impact variables ($J_b$, $M_d$, $F_d^{Fe}$, and $\delta_{pd}$). Notably, it illustrates the post-impact parameter spaces sampled by prior studies, which represent different impact scenarios. These scenarios include the canonical giant impact scenario \citep{canupSimulationsLateLunarforming2004,canupLunarformingCollisionsPreimpact2008}, equal-mass mergers \citep{canupFormingMoonEarthlike2012}, and the fast spinning proto-Earth scenario \citep{cukMakingMoonFastSpinning2012}. We note that our simulations produce similar outcomes to each of these studies, with the exception of the region $\SI{2.5}{\JEM}\leq J_b\leq\SI{2.75}{\JEM}$, where \citet{canupFormingMoonEarthlike2012} reports slightly larger disk masses. However, since our study covers a much larger region of the pre-impact parameter space and does not focus on a specific impact scenario, our collision outcomes are more diverse.

\subsubsection{Disk mass}
\begin{figure*}[ht!]
\centering
\plotone{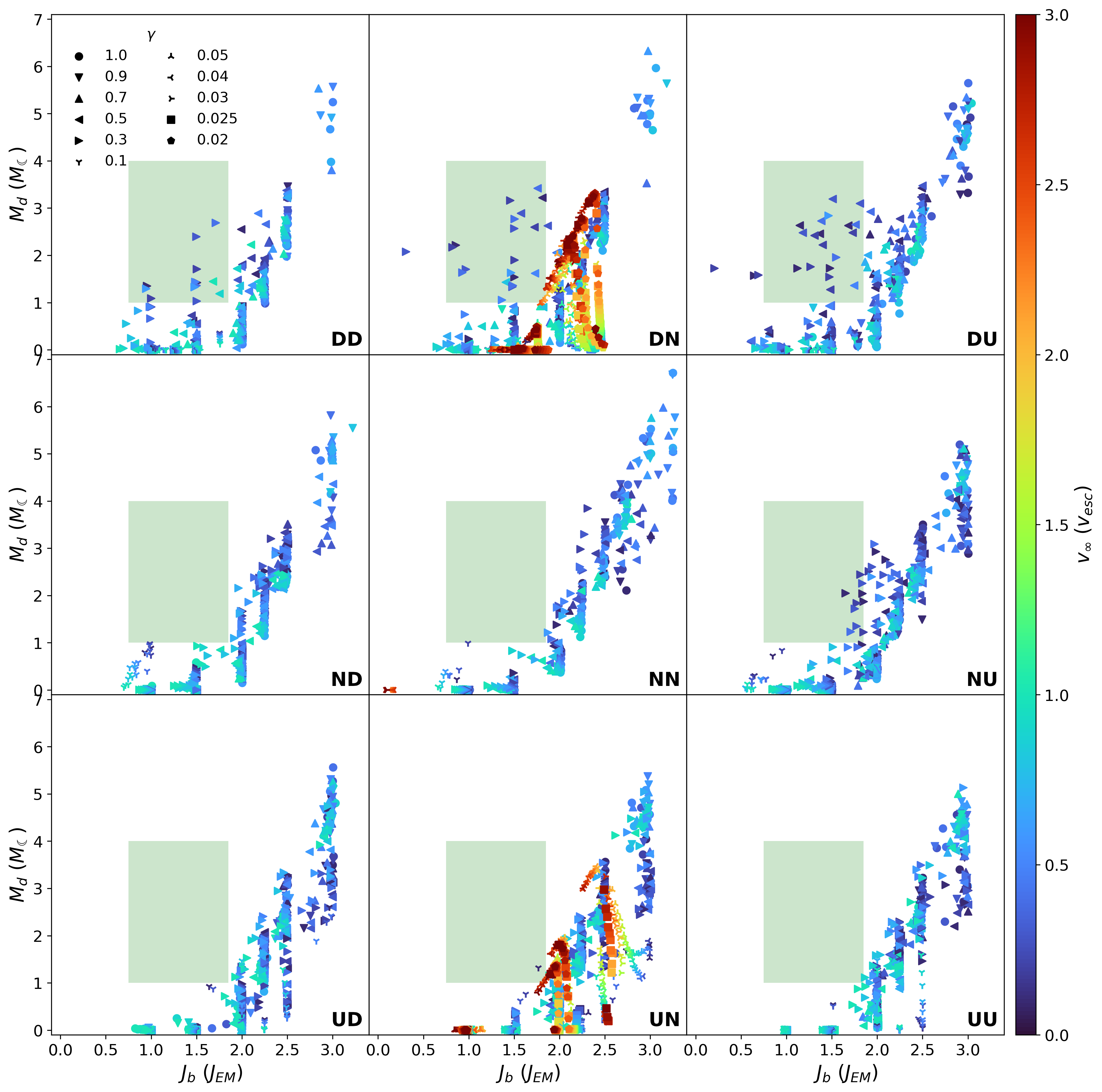}
\caption{Relation between disk mass $M_d$ and bound angular momentum $J_b$ for all 6247 analyzed simulations. This figure shows the same data as the top panel of Figure~\ref{fig:all_other_paper_data_lines} but divided up into the nine different spin orientations described in Section~\ref{sec:initial_conditions}. In Paper I we find that in order to create a massive disk ($M_d \geq \SI{2}{\Mmoon}$) with non-rotating bodies, an initial angular momentum of $J_0 \geq \SI{2.25}{\JEM}$ is needed. We also find that impacts with impactor-to-target mass ratios $\gamma < 0.1$ do not result in significant disks. The shaded green box shows the region of interest that is not sampled by NN cases ($\SI{0.75}{\JEM} \leq J_b \leq \SI{1.85}{\JEM}$ and $\SI{1.0}{\Mmoon} \leq M_d \leq \SI{4.0}{\Mmoon}$). Adding pre-impact rotation to the colliding bodies, we find massive disks in this region and particularly with M$_{d} \geq \SI{2}{\Mmoon}$ for J$_b \sim \SI{1}{\JEM}$. These disks are produced by initial conditions with counter-rotating targets (DX). Furthermore, high-velocity, low-$\gamma$ impacts (red dots) can produce massive disks for rotating targets but result in excess bound AM.}
\label{fig:DiskMass_vs_J_bound_configurations}
\end{figure*}

\begin{figure*}[ht!]
\centering
\plotone{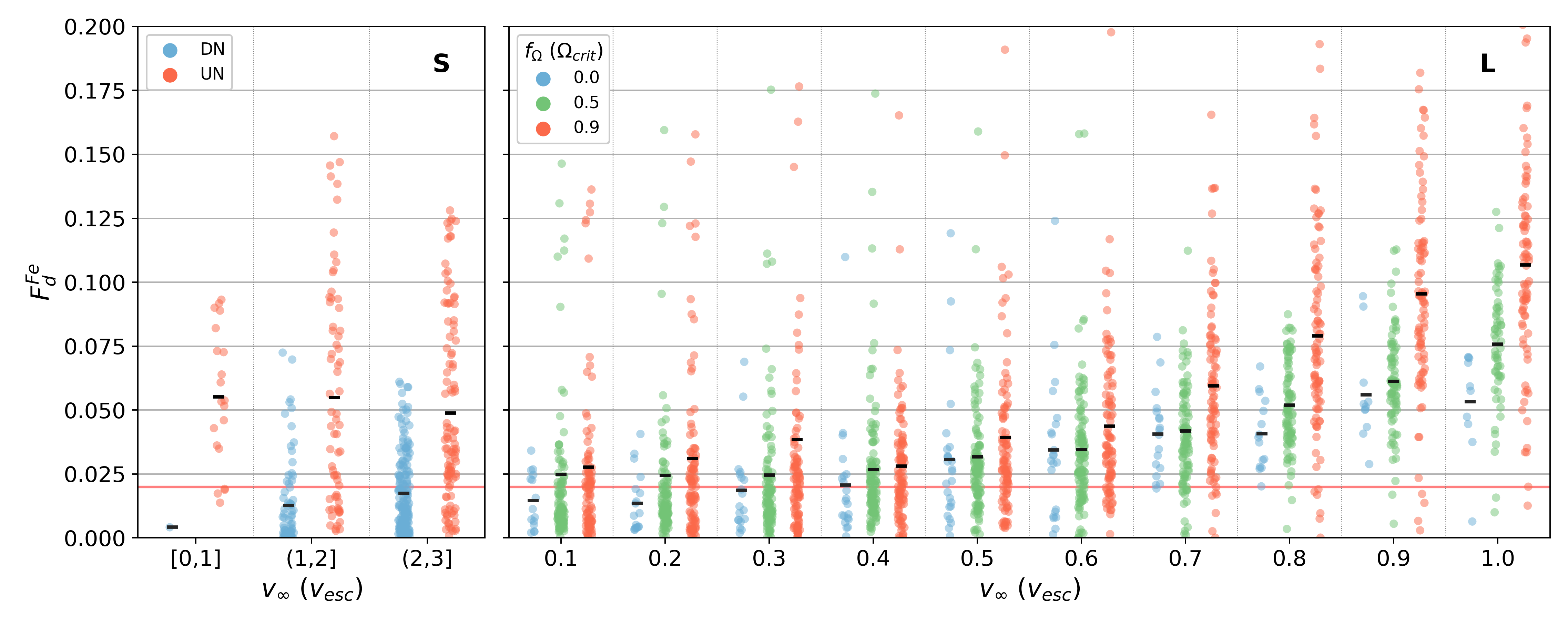}
\caption{The iron mass fraction of the protolunar disk ($F^{Fe}_d$) for collisions that generate a disk of at least one lunar mass ($M_d \geq \SI{}{\Mmoon}$). Generally, it is difficult to place a substantial amount of iron in orbit and massive disks therefore tend to be depleted in iron. In collisions between non-rotating bodies, $F^{Fe}_d$ evinces a positive correlation with $v_{\infty}$, with higher velocities tending to produce larger disk iron mass fractions. In collisions where at least one body is rotating, this dependence on $v_{\infty}$ persists for large impactors ($\gamma \geq 0.1$; Panel L) and an additional dependence on the rate of rotation ($f_{\Omega}$) can be seen. For small impactors ($\gamma < 0.1$; Panel S), impacts on a counter-rotating target (DN) tend to produce significantly lower disk iron mass fractions than impacts on a co-rotating target (UN). Thus, slow impacts by non-rotating or slowly rotating bodies are preferred to reproduce the disk iron mass fraction constraint.} 
\label{fig:Fe_disk}
\end{figure*}

For non-rotating collisions, the mass of the generated circumplanetary disk ($M_d$) is mostly determined by the pre-impact angular momentum budget ($J_0$), evincing a correlation of $r=0.89$. Moreover, collisions between non-rotating bodies require a pre-impact angular momentum budget of at least twice the current Earth-Moon angular momentum budget ($J_0 \geq \SI{2}{\JEM}$) to produce a disk of at least one lunar mass ($M_d \geq \SI{}{\Mmoon}$). Notably, this implies that the canonical Moon-forming impact model can not form sufficiently massive disks. 

In Figure~\ref{fig:DiskMass_vs_J_bound_configurations}, we show that the strong correlation between $J_0$ and $M_d$ persists for collisions between rotating bodies. Indeed, $J_0$ and $M_d$ are strongly correlated at $r=0.80$, while $J_b$ and $M_d$ are also strongly correlated at $r=0.78$. Despite this persisting correlation, we demonstrate that impacts with a counter-rotating target (DD, DN, DU) can produce significantly more massive disks at lower values of $J_b$. This population of ''low-$J_b$, high-$M_b$'' outcomes can be seen in Figure~\ref{fig:all_other_paper_data_lines} (top panel), while Figure~\ref{fig:DiskMass_vs_J_bound_configurations} clearly demonstrates that these outcomes are unique to the DX configurations. Furthermore, Figure~\ref{fig:DiskMass_vs_J_bound_configurations} shows that this population of collisions is limited to moderate mass ratios ($0.3 \leq \gamma \leq 0.5$). The NU configuration also shows hints of a similar population, but does not achieve similarly low values of $J_b$ as seen in the DX cases.

The population of low-$J_b$, high-$M_d$ outcomes is produced only by grazing ($\theta_{imp} \gtrsim \SI{45}{\degree}$), low velocity ($v_{\infty} \lesssim 0.5~v_{esc}$) impacts onto counter-rotating targets (DX) at moderate mass ratios ($0.3 \leq \gamma \leq 0.5$). We note that several collisions in this population satisfy the disk iron mass fraction constraint, however none of the collisions produce planets and disks with similar compositions. Because the post-impact planets in this population are rotating well below the HSSL, they are therefore not candidates for post-impact compositional equilibration of the planet and disk (e.g., via a synestia). \emph{This implies that, for the collisions in this population to be viable Moon-forming impacts, their targets and impactors would have to evince very similar isotopic compositions prior to impact.} This population represents a novel class of Moon-forming impacts, wherein a counter-rotating target roughly the mass of Venus suffers a grazing, low-velocity impact by an impactor roughly 2-3 times the mass of Mars.

Generally, for a given value of $J_b$, counter-rotating targets (DD, DN, DU) are capable of generating larger disk masses than non-rotating targets (ND, NN, NU) while co-rotating targets (UD, UN, UU) tend to result in even smaller disk masses. This is because, for co-rotating targets, the material near the contact zone with the impactor is moving in the same direction as the impactor itself, effectively swallowing it. In the case of counter-rotating targets, this effect is reversed, meaning that the local material is moving towards the impactor, thereby producing larger \emph{local} collision velocities. The higher local relative velocity at the impact site could result in high vapor mass fraction of the disk. Future work should investigate this using a higher number of particles and a more sophisticated EOS to connect to studies investigating lunar accretion \citep{nakajimaLimitedRoleStreaming2024}.

For large impactors ($\gamma \geq 0.1$), the rotation rate ($f_\Omega$) of the colliding bodies affects the disk mass. For counter-rotating (DX) and non-rotating targets (ND and NU), faster rotation rates result in lower disk masses. For co-rotating targets (UX), the relationship is reversed; indeed, while the UD configuration does not show a dependence on $f_\Omega$, faster rotation rates result in more massive disks for both the UN and UU configurations.

In Paper I, we found that small impactors ($\gamma < 0.1$) between non-rotating bodies (NN) result in very small disk masses. However, for small impactors with rotating targets, (DN and UN), such low-$\gamma $ collisions are able to produce disks with significant mass for $f_\Omega=1.01$. The $f_\Omega=1.01$ cases result in significantly more massive disks than $f_\Omega=0.9$ (for which there are no cases with $M_d\geq\SI{2.0}{\Mmoon}$) and $f_\Omega=0.5$ (which does not produce significant disk mass at all). This implies that, for rotating bodies, it is much easier to eject mass because the material at the equator is more weakly bound than in the non-rotating case. Furthermore, we confirm the results of \citet{cukMakingMoonFastSpinning2012} that collisions with low $\gamma$ and high $v_{\infty}$ are able to generate massive disks but result in excess angular momentum of the Earth-Moon system ($J_{b} \geq \SI{2}{\JEM}$).

In summary, to satisfy the disk mass constraint ($M_d > \SI{}{\Mmoon}$), two results are useful to take note of. First, with pre-impact rotation, the disk mass remains strongly correlated with the post-impact angular momentum budget ($J_b$), with $J_b \geq\SI{2}{\JEM}$ generally required to produce sufficiently massive disks. Second, the tyranny of this relationship can be broken by a population of grazing, low-velocity collisions in which a medium-size impactor ($0.3 \leq \gamma \leq 0.5$) strikes a counter-rotating target (DX). This population can produce massive disks ($M_d > \SI{2}{\Mmoon}$) with low iron mass fractions ($F^{Fe}_d \leq 0.02$), but cannot meet the composition constraint.

\subsubsection{Disk iron mass fraction}
For non-rotating collisions, we showed that $F^{Fe}_d$ is most strongly correlated with $v_{\infty}$ ($r=0.48$), with high-velocity impacts tending to produce disks more enriched in iron. However, for collisions with pre-impact rotation this is not the case ($F^{Fe}_d$ shows but a weak correlation of $r=0.17$ with $b_{\infty}$).

In the collisions presented here, $F_d^{Fe}$ spans the entire range of possible compositions, from pure rock ($F_d^{Fe}=0$) to pure iron ($F_d^{Fe} = 1$). While there is no clear relation between $F_d^{Fe}$ and $J_{b}$ or $M_{d}$, there is a general trend that the maximum iron mass fraction decreases with higher $M_d$ and $J_b$. Extremely iron rich disks can be obtained up to $J_b \sim \SI{2}{\JEM}$ and tend to have a very low disk mass. A similar trend is observed for the lowest possible iron mass fraction that decreases with higher $M_d$ and $J_b$. Pure rock disks are found up to \SI{3}{\Mmoon}. The minimum iron mass fraction decreases for higher mass disks because the contribution from a single iron particle is lower. Increasing the resolution of the simulation would allow to resolve lower particle masses and therefore lower iron mass fractions.

For collisions between non-rotating bodies (NN), $F^{Fe}_d$ is positively correlated with $v_{\infty}$. In Figure \ref{fig:Fe_disk}, for collisions between rotating bodies, we show that this dependence on $v_{\infty}$ persists for large impactors ($\gamma \geq 0.1$). In addition, $F^{Fe}_d$ is positively correlated with the pre-impact rotation rate of the rotating bodies ($f_{\Omega}$). For small impactors ($\gamma < 0.1$) on rotating targets, $F^{Fe}_d$ does not show the same dependence on $v_{\infty}$ or $f_{\Omega}$. Instead, $F^{Fe}_d$ is strongly affected by the direction of the target's rotation, with impacts on co-rotating targets (UN) producing much higher disk iron mass fractions.

To satisfy the disk iron mass fraction constraint ($F^{Fe}_d \leq 0.02$), the following systematic trends can be ascertained from Figure \ref{fig:Fe_disk}. For large impactors ($\gamma \geq 0.1$), relatively low-velocity impacts between non-rotating or slowly rotating bodies are preferred. For small impactors ($\gamma < 0.1$), impacts with a counter-rotating target (DN) produced significantly smaller disk iron mass fractions and are therefore preferred over impacts with a co-rotating target (UN).

\subsubsection{Planet-disk compositional difference}
\begin{figure}[ht!]
\centering
\plotone{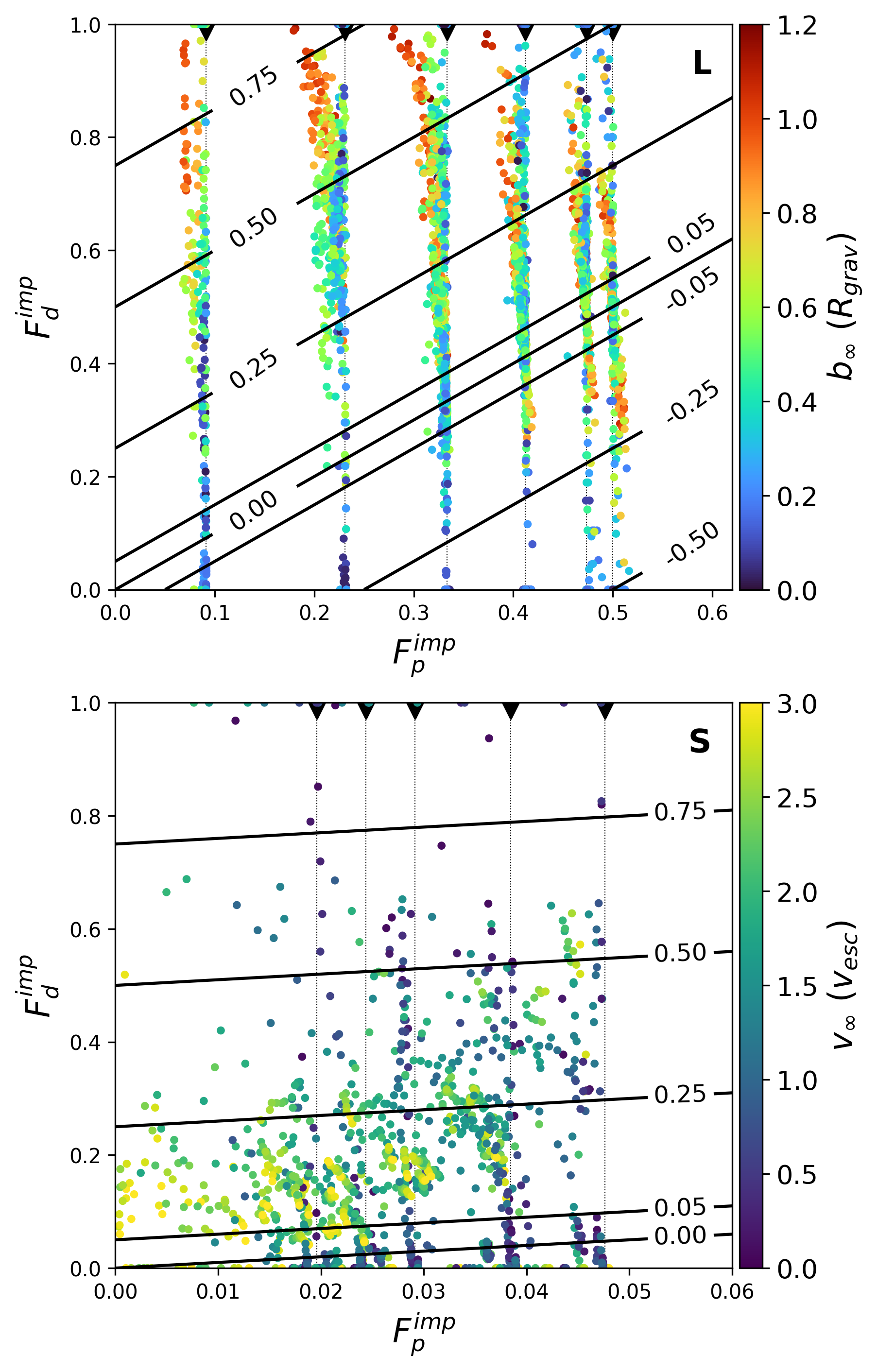}
\caption{The mass fractions of impactor material in the proto-Earth ($F^{imp}_p$) and protolunar disk ($F^{imp}_d$) following the impact. The compositional similarity of the proto-Earth and disk ($\delta_{pd}$) is determined by the difference in these mass fractions, as defined in Equation \ref{eq:delta-mix}. Isolevels for $\delta_{pd}$ are shown by the diagonal lines. The vertical lines indicate values of $\gamma$. For large impactors (Panel L; $\gamma \geq 0.1$), the impactor mass fraction of the proto-Earth ($F^{imp}_{p}$) is almost entirely determined by $\gamma$; this is because---for the range of impacts considered here---the impactor generally merges almost entirely with the target, with only a relatively small fraction contributing to the disk. For small impactors ($\gamma < 0.1$), this relationship only holds weakly for low-velocity ($v_{\infty} \leq v_{esc}$) impacts. For higher-velocity impacts ($v_{\infty} > v_{esc}$), $F^{imp}_{p}$ is no longer correlated with $\gamma$.}
\label{fig:mixing_parameter}
\end{figure}

\begin{figure*}[ht!]
\centering
\plotone{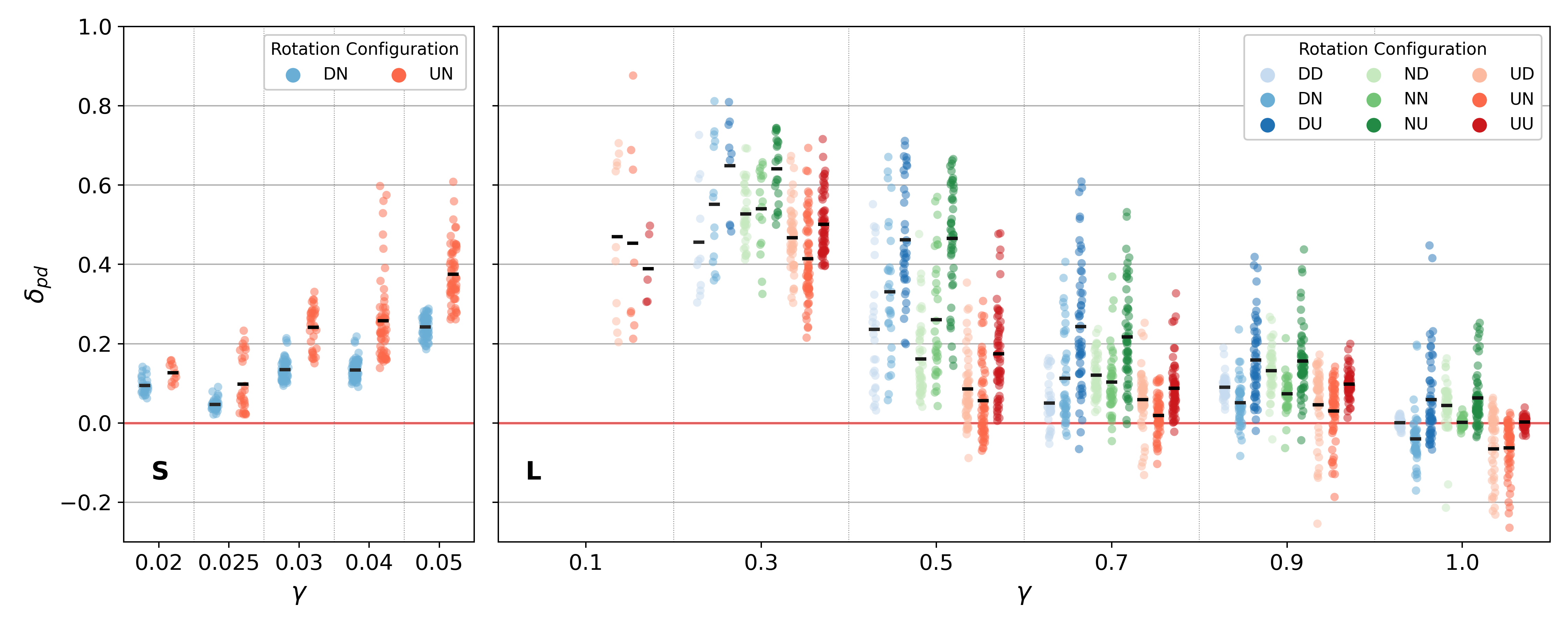}
\caption{Effect of rotation on the planet-disk compositional difference ($\delta_{pd}$) for collisions that generate a disk of at least one lunar mass ($M_d \geq \SI{}{\Mmoon}$). For large impactors (Panel L; $\gamma \geq 0.1$), $\delta_{pd}$ is most strongly correlated with $\gamma$, with higher mass ratios producing planets and disks that are more similar in composition. Additionally, for large impactors, $\delta_{pd}$ is also influenced by the rotation of the impactor, whereby co-rotating impactors (XU) tend to produce disks significantly more enriched in impactor material relative to the planet. For small impactors (Panel S; $\gamma < 0.1$), $\delta_{pd}$ is also most strongly correlated with $\gamma$, albeit with lower mass ratios producing planets and disks more similar in composition. In contrast to large impactors, $\delta_{pd}$ for small impactors depends on the rotation of the target, with impacts on co-rotating targets (UN) producing disks more enriched in impactor material relative to the planet. Very low compositional differences require either very small or very large impactors as proposed in prior work \citep{cukMakingMoonFastSpinning2012, canupFormingMoonEarthlike2012}.}
\label{fig:delta_pd_rotation_dependence}
\end{figure*}

The compositional difference we use as the proxy for the isotopic similarity is defined as

\begin{equation} \label{eq:delta-mix}
    \delta_{pd} = F_d^{imp} - F_p^{imp} = \left(\frac{N_{imp}}{N_{tot}}\right)_{d} - \left(\frac{N_{imp}}{N_{tot}}\right)_{p} \,,
\end{equation}

\noindent where $N_{imp}$ is the number of particles originating from the impactor and $N_{tot}$ is the total number of particles in either the disk ($d$) or the planet ($p$). It exhibits a degeneracy, where different disk compositions $F_d^{imp}$ can result in the same value for $\delta_{pd}$ depending on the corresponding planet composition. This is shown in Figure~\ref{fig:mixing_parameter} together with all data points.

The composition $\delta_{pd}$ varies from $\delta_{pd} \sim -0.5$, where the disk material has a lower impactor fraction than the planet, to being formed purely of impactor material ($\delta_{pd} \sim 1$). None of the disks, no matter how low the mass, are composed entirely of target material. Disks entirely formed from the impactor are possible but tend to be low in mass. Such disks can also have a high iron mass fraction. As the bound angular momentum and disk mass increase, the composition tends to be more well mixed and the most massive disks have impactor mass fractions very similar to the planet. Those disks are generated in near-equal mass collisions which is consistent with \citet{chauCouldUranusNeptune2021}.

In Paper I, we found that only near equal mass mergers (i.e., $\gamma \sim 1$) can achieve near perfect mixing ($\delta_{pd} \sim 0$) in the absence of pre-impact rotation. This appears to be a result of the symmetry of the impact. Once pre-impact rotation is introduced, many cases with perfect mixing are still from $\gamma = 1$ impacts (435 of 1315 with $\left|\delta_{pd}\right|\leq 0.05$). However, including pre-impact rotation can break the symmetry that would otherwise exist without pre-impact rotation. In these cases, the body that is co-rotating with the collision (i.e., in the ``up'' configuration) tends to have a higher mass fraction in the disk than in the planet. We also obtain cases with $\left| \delta_{pd} \right| \leq 0.01$ for lower impactor-to-target mass ratios. If we additionally require the disk mass to be at least \SI{1}{\Mmoon} then $\gamma \geq 0.5$ impacts can still result in perfectly mixed disks. The impacts with $\gamma = 0.5$ are all involving a co-rotating target which enhances the ejection of material originating from the proto-Earth. The collisions with $\gamma < 0.1$ proposed by \citet{cukMakingMoonFastSpinning2012} can also produce near perfect mixing ($\left| \delta_{pd} \right| \leq 0.01$), but in these cases the disk mass is too small to allow the formation of the Moon.

From Figures \ref{fig:mixing_parameter} and \ref{fig:delta_pd_rotation_dependence}, several useful trends can be extracted. First, for large impactors, the impactor mass fraction of the proto-Earth ($F^{imp}_p$) is determined almost exactly by the pre-impact mass ratio ($\gamma$) of the colliding bodies. This is because, for the impacts in the large impactor set, the impactors tends to merge almost entirely with their targets. In contrast, the impactor mass fraction of the disk ($F^{imp}_{d}$) does not show any clear dependencies, but we note that planet and disks with similar compositions tend to result from more head-on impacts. Furthermore, for disks of at least one lunar mass ($M_d \geq \SI{}{\Mmoon}$), only high-$\gamma$ collisions can produce sufficiently low values of $F^{imp}_{d}$ to achieve favorable values of $\delta_{pd}$. For small impactors, the dependence of $F^{imp}_p$ on $\gamma$ only holds weakly for low-velocity impacts ($v_{\infty} \leq v_{esc}$). For high-velocity impacts ($v_{\infty} > v_{esc}$) by small impactors, $\gamma$ can no longer be used to predict $F^{imp}_p$, however $F^{imp}_d$ shows a dependence on $\gamma$ in this region. Because only the high-velocity impacts are capable of producing disks of at least one lunar mass; for these impacts, only the lowest values of $\gamma$ can produce favorable values of $\delta_{pd}$.

To satisfy the planet-disk compositional similarity constraint for massive disks ($M_d \geq \SI{}{\Mmoon}$), several systematic trends can be leveraged. For large impactors, only high-$\gamma$ impacts $(\gamma \geq 0.5)$ are capable of producing favorable compositions ($\left\vert\delta_{pd}\right\vert \leq 0.05$). Within this region, impacts by non-rotating (XN) or counter-rotating impactors (XD) are likely to produce smaller compositional differences. For small impactors, only very small mass ratios of $\gamma \leq 0.025$ and velocities in excess of $2~v_{esc}$ are able to produce disks of at least one lunar mass with favorable compositions; additionally, impacts onto a counter-rotating target (DN) tend to produce lower compositional differences than impacts onto co-rotating targets (UN).

\subsubsection{Hot-spin stability limit (HSSL)}
\begin{figure}[ht!]
\centering
\plotone{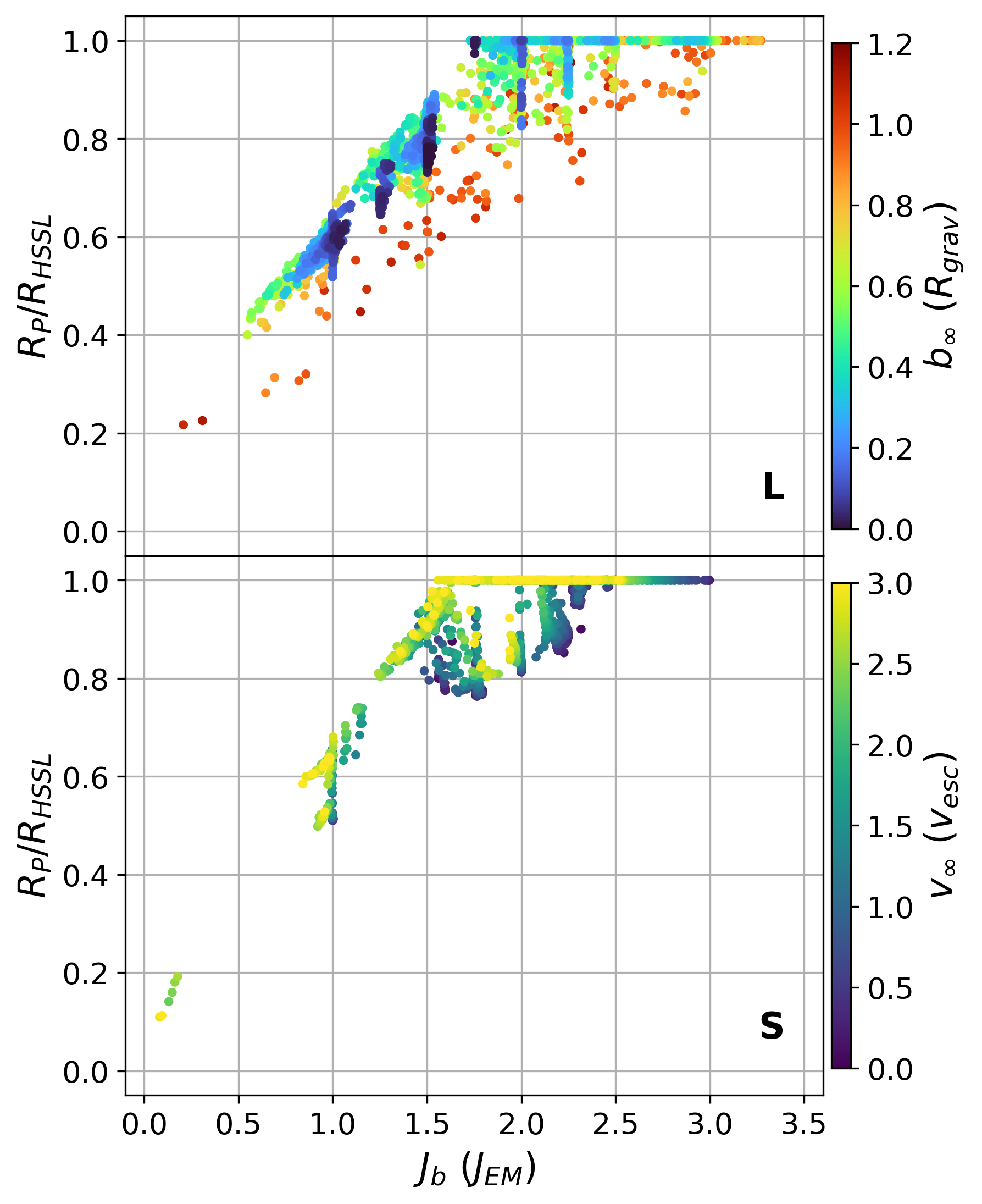}
\caption{Proximity to the hot-spin stability limit (HSSL) following an impact. The post-impact state of the collision is considered to be at or above HSSL if the equatorial radius of the proto-Earth is equal to the rotational stability limit ($R_p/R_{HSSL} = 1$). Only collisions that are at or above the HSSL can continue to exchange material between the proto-Earth and protolunar disk, thereby allowing post-impact processes to equilibrate their compositions. For large impactors (Panel L; $\gamma \geq 0.1$), a population of grazing, low-velocity, low-$\gamma$ collisions tends to be farther from the HSSL. For small impactors (Panel S; $\gamma < 0.1$), this population shows only a dependence on $v_{\infty}$, with lower-velocity impacts being farther from the HSSL.}
\label{fig:HSSL_vs_Jb}
\end{figure}

If the atmosphere of the proto-Earth and the inner edge of the protolunar disk remain in contact following the impact, then it is possible for these reservoirs to continue exchanging material. The resulting post-impact structure is known as a synestia \citep{lockOriginMoonTerrestrial2018} and may allow the Earth and protolunar disk to achieve near or total compositional equilibrium, potentially relaxing the isotopic constraints. However, for a synestia to exist, the post-impact Earth must be rotating at a rate sufficient to push its equatorial radius to the hot-spin stability limit ($R_p/R_{HSSL} = 1$). $R_{HSSL}$ is determined by the disk finder; a detailed explanation of how $R_{HSSL}$ is calculated can be found in Appendix B of Paper I.

For collisions without pre-impact rotation, we demonstrated that the post-impact Earth's proximity to $R_{HSSL}$ is largely determined by the pre-impact angular momentum budget ($J_0$). For $\gamma \gtrsim 0.6$, $J_0$ appears to be the sole determining variable. For $\gamma \lesssim 0.6$, the impact velocity ($v_{\infty}$) also plays a small role, with higher impact velocities resulting in decreased proximity to the HSSL (i.e., lower values of $R_p/R_{HSSL}$) for a given pre-impact angular momentum budget. For all non-rotating collisions, a pre-impact angular momentum budget of $J_0 \geq \SI{2}{\JEM}$ is required to reach the HSSL. For non-rotating collisions, this implies that any collision with a pre-impact angular momentum budget of $J_0 \leq \SI{2}{\JEM}$ cannot invoke post-impact compositional mixing.

For collisions between rotating bodies, the strong dependence on the pre-impact angular momentum budget ($J_0$) persists. Figure \ref{fig:HSSL_vs_Jb} demonstrates this relationship for both large ($\gamma \geq 0.1$) and small impactors ($\gamma < 0.1$). Whereas large impactors appear to reach the HSSL only at $\SI{1.75}{\JEM}$, small impactors can reach the HSSL as soon as $\SI{1.5}{\JEM}$. For large impactors, a significant number of grazing, low-velocity, low-$\gamma$ impacts populate an area below the otherwise well-behaved relation. We note that this phenomenon was also present in Paper I for collisions between non-rotating bodies, whereby low-velocity, low-$\gamma$ impacts can be seen receding from the HSSL as the impact angle becomes large due to the increasing angular momentum budget. For small impactors, there also exists a population of points below the otherwise well-behaved relation. However, unlike for large impactors, this population does not evince an obvious dependence on $\gamma$ or the impact angle. In Figure \ref{fig:HSSL_vs_Jb} (Panel S), a dependence on $v_{\infty}$ can be discerned, with lower-velocity impacts being furthest from the HSSL.

\subsection{Promising cases}\label{sec:Promising_Cases}
\begin{figure*}[ht!]
\centering
\plotone{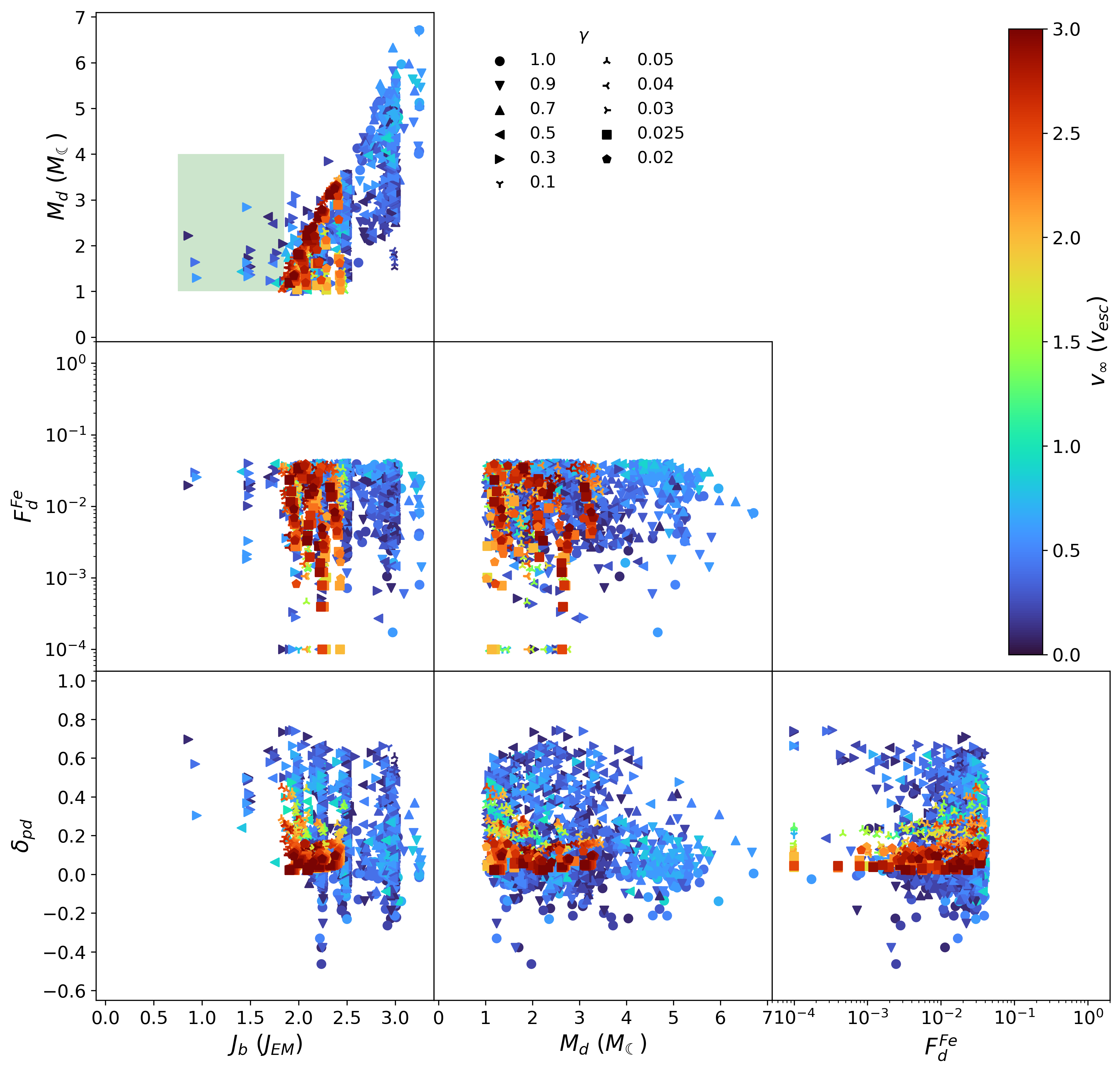}
\caption{All 1570 cases that remain after applying the permissive constraints ($M_b\geq \SI{1}{\Mearth}$, $M_d \geq \SI{1}{\Mmoon}$ and $F_d^{Fe}\leq \SI{0.04}{}$) to all results. Disk iron mass fractions of exactly zero are moved to \SI{1e-4}{} to be visible on a logarithmic plot (for details see caption of Figure~\ref{fig:Seaborn_Triangle_plot_only_output_variables}). This subset contains collisions from all rotation configurations, all impactor-to-target mass ratios $\gamma$ and all sampled velocities $v_{\infty}$. While most results are beyond $J_b\geq\SI{1.5}{\JEM}$, there are three results at $J_b\simeq\SI{1}{\JEM}$ but those have disks with a higher impactor fraction than the planet. Cases with good mixing appear at $J_b\geq\SI{1.75}{\JEM}$.}
\label{fig:successful_cases_loose}
\end{figure*}

\begin{figure*}[ht!]
\centering
\plotone{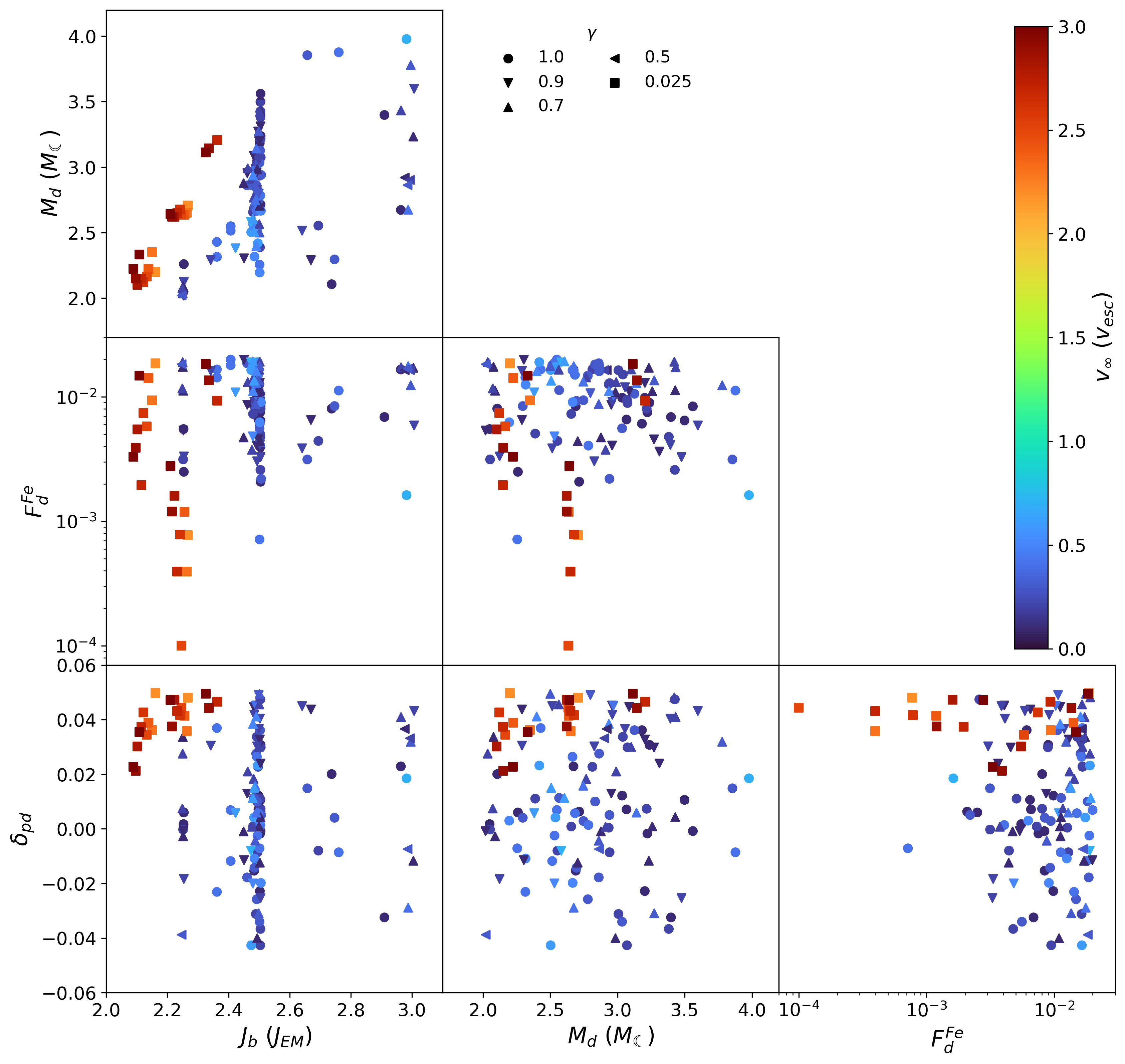}
\caption{All 135 cases that remain after applying the strict constraints ($M_b\geq \SI{1}{\Mearth}$, $\SI{2}{\Mmoon} \leq M_d \leq \SI{4}{\Mmoon}$, $F_d^{Fe}\leq \SI{0.02}{}$ and $\left\vert\delta_{pd}\right\vert\leq \SI{0.05}{}$). Disk iron mass fractions of exactly zero are moved to \SI{1e-4}{} to be visible on a logarithmic plot. (for details see caption of Figure \ref{fig:Seaborn_Triangle_plot_only_output_variables}). Note the change in axis limits compared to Figure~\ref{fig:successful_cases_loose}. Under these constraints, there are no results around $J_b\simeq\SI{1}{\JEM}$ and the minimum bound angular momentum is $J_b=\SI{2.09}{\JEM}$. Only the higher impactor-to-target mass ratios $\gamma\geq0.5$ produce promising cases, with the exception of $\gamma=0.025$ which confirms the findings of \citet{cukMakingMoonFastSpinning2012}.}
\label{fig:successful_cases_tight}
\end{figure*}

A successful Moon-forming impact must satisfy a set of constraints from observations, measurements, and theory. Most of these constraints are drawn from measurements of the Earth-Moon system as it exists today, while others are provided by theoretical studies of the system's past evolution. In the context of giant impact simulations, these observations translate into constraints on the simulated post-impact properties of the proto-Earth and the protolunar disk (for a detailed discussion of these properties see Section 2 of Paper I). These constraints are the mass of the Earth for which we use the total mass of the gravitationally bound material ($M_b$), the total angular momentum of the bound material ($J_b$), the circumplanetary disk mass ($M_d$), the iron mass fraction of the disk ($F_d^{Fe}$), and the relative fraction of impactor material in the disk relative to the proto-Earth ($\delta_{pd}$). The Earth also has a well known iron mass fraction, but we ignore this constraint because we set the initial core fractions of both the target and the impactor to \SI{0.33}{} and the total mass to \SI{1.05}{\Mearth}, such that for results that satisfy the constraint on $M_b$, the resulting iron mass fraction of the planet is satisfying the constraint.

Despite the addition of several thousand new simulations that include pre-impact rotation to the data set presented in Paper I, we are still unable to identify a single impact scenario that can simultaneously satisfy all known constraints. Specifically, a collision that generates a sufficiently massive protolunar disk ($M_d \geq \SI{2}{\Mmoon}$) and recovers the current angular momentum of the Earth-Moon system ($J_b = \SI{}{\JEM}$) along with good compositional mixing remains elusive. But we can apply a subset of constraints on the data to find promising cases. In Paper I, we considered two different sets of constraints: one permissive ($M_b\geq\SI{1}{\Mearth}$, $M_d\geq\SI{1}{\Mmoon}$ and $F_d^{Fe}\leq 0.04$) and one strict ($M_b\geq\SI{1}{\Mearth}$, $\SI{2}{\Mmoon} \leq M_d \leq \SI{4}{\Mmoon}$, $F_d^{Fe}\leq \SI{0.02}{}$ and $\left\vert\delta_{pd}\right\vert\leq \SI{0.05}{}$). In both cases there is no constraint on the bound angular momentum because for the non-rotating bodies in Paper I we do not obtain massive disks ($M_d\geq \SI{1}{\Mmoon}$) below $J_b \sim \SI{2}{\JEM}$.

In Figure~\ref{fig:successful_cases_loose}, the 1570 promising cases under the assumption of the permissive constraints are shown. This subset of results contains cases from all rotation configurations, all impactor-to-target mass ratios $\gamma$ and all sampled velocities $v_{\infty}$. From the cases with counter-rotating target (i.e., DX configurations), we get some results around $J_b\simeq\SI{1}{\JEM}$, but most of the results are beyond $J_b\geq\SI{1.5}{\JEM}$. Cases with good mixing start to appear at $J_b\geq\SI{1.75}{\JEM}$. Of the cases populating the green region of Figure~\ref{fig:DiskMass_vs_J_bound_configurations}, 28 satisfy these permissive constraints.

Figure~\ref{fig:successful_cases_tight} shows the 135 promising cases under the strict constraints. Under these constraints, we can no longer reconcile the bound angular momentum and disk mass, which confirms the findings of Paper I, even though pre-impact rotation is considered. Like the subset with permissive constraints, this subset also contains promising cases of all rotation configurations. In general, either higher $\gamma\geq0.5$ or very low ($\gamma=0.025$, as proposed by \citealt{cukMakingMoonFastSpinning2012}) impactor-to-target mass ratios are able to generate promising cases. For $\gamma\geq0.5$ all promising cases are with asymptotic relative velocities of $v_{\infty}\leq0.7\,v_{esc}$, while for $\gamma=0.025$, all promising cases are for $2.2\,v_{esc}\leq v_{\infty}\leq 3.0\, v_{esc}$. The minimum bound angular momentum in this subset is $J_b=\SI{2.09}{\JEM}$. Of the cases populating the green region of Figure~\ref{fig:DiskMass_vs_J_bound_configurations}, none satisfy these strict constraints because they all have disk compositions that are more enriched in impactor material than the planet and thus do not satisfy the constraint on $\left|\delta_{pd}\right|$.

In principle, the cases that do not result in a large enough bound mass, i.e., $M_b\geq\SI{1.0}{\Mearth}$, could be promising if the initial mass of the proto-Earth was slightly larger prior to the collision. We therefore also investigate how allowing for a lower bound mass affects the above findings. For the permissive constraints (see Figure \ref{fig:successful_cases_loose}) we find that relaxing the constraint on the bound mass to $M_b\geq \SI{0.95}{\Mearth}$ increases the number of promising cases from 1570 to 1633. All of the additional cases are from low-$\gamma$, high velocity impacts with $J_b > \SI{1.5}{\JEM}$. For the strict constraints (Figure \ref{fig:successful_cases_tight}) no additional promising simulations show up.

Of all the known constraints, $F_d^{Fe}$ is the easiest to satisfy. This is due to the fact that, in general, it is difficult to inject significant amount of iron into the disk. Therefore, we focus on the relationship between the remaining three constraints ($M_d$, $J_b$ and $\left\vert\delta_{pd}\right\vert$). If we apply a constraint on the bound angular momentum ($\SI{0.75}{\JEM}\leq J_b\leq \SI{1.25}{\JEM}$) together with the strict constraints we can not find results, as discussed above. But from these three constraints we can choose any combination of two and we get results that satisfy them.

\subsection{Immediate satellite formation}\label{sec:immediate_formation}
\begin{figure*}[ht!]
\centering
\plotone{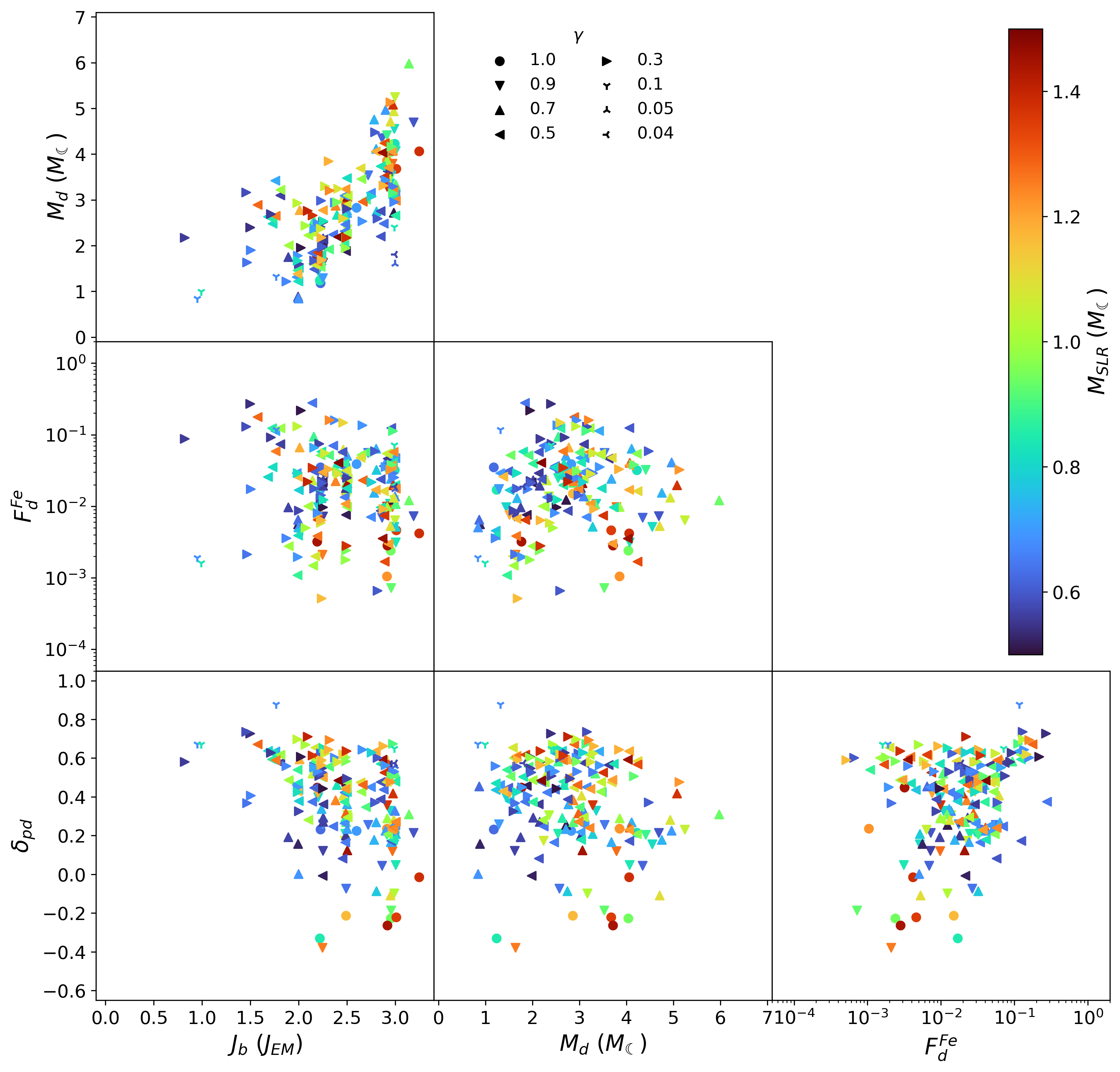}
\caption{All 191 cases that produce a bound second largest remnant (SLR) with a mass of $\SI{0.5}{\Mmoon} \leq M_{SLR} \leq \SI{1.5}{\Mmoon}$ at the end of the simulation. The results are very diverse, spanning nearly the whole post-impact parameter space. While none of these simulations are consistent with the strict constraints, 137 satisfy the permissive constraints (see Section~\ref{sec:immediate_formation} for details).}
\label{fig:considerable_M_SLR_Disk_Mass}
\end{figure*}

\begin{figure}[ht!]
\centering
\plotone{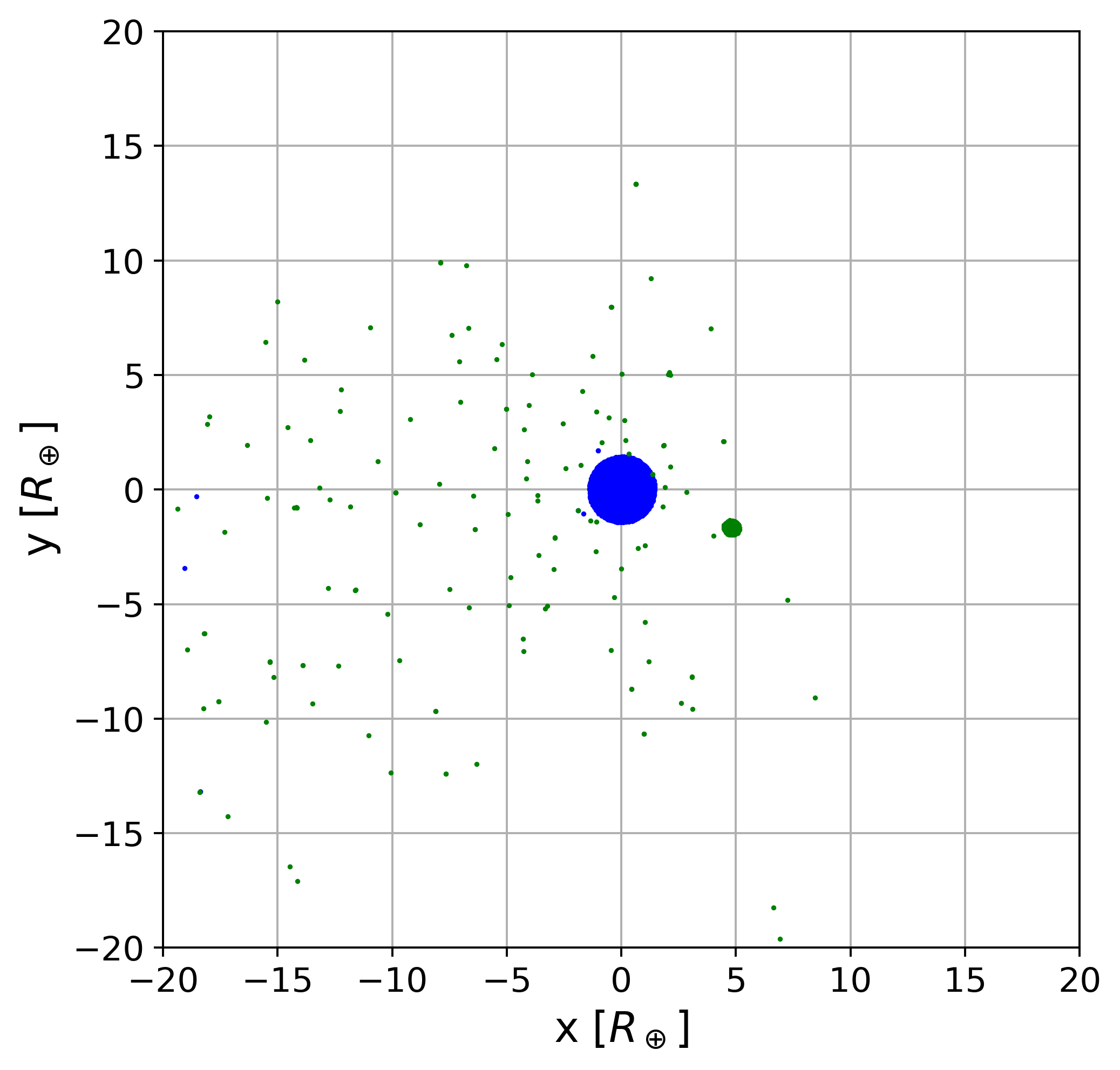}
\caption{Plot of the end state of run 3567 (for pre-impact and post-impact parameters see Section~\ref{sec:immediate_formation}). Depicted is a projection of the particles into the plane of the collision. Blue particles belong to the planet while green particles form the disk. This run results in a massive bound second largest fragment with $M_{SLR}=\SI{0.84}{\Mmoon}$ in a disk of $M_d=\SI{0.99}{\Mmoon}$ at a bound angular momentum very close to the Earth-Moon system $J_b=\SI{0.99}{\JEM}$. The semi-major axis of the fragments orbit is \SI{7.2}{\Rearth} and it has an eccentricity of $e=0.3376$.}
\label{fig:NN_J1000_G0100_V02000_B08153_OMEGA0000}
\end{figure}

In several cases, we find gravitationally bound fragments that remain on stable orbits until the end of the simulation. Some of these fragments have masses around \SI{1}{\Mmoon} and may be considered immediately formed Moons as proposed in \citet{kegerreisImmediateOriginMoon2022} (and earlier by \citet{chauCouldUranusNeptune2021} in the context of Uranus and Neptune). Such (potential) satellites are either formed directly from the tidal disruption of the impactor (and therefore have a higher impactor fraction than the planet) or due to the fragmentation of spiral arms (which results in a well mixed composition or even disks with a lower impactor fraction than the planet). Additionally, \citet{kegerreisImmediateOriginMoon2022} show that such objects can accrete a substantial amount of material from the circumplanetary disk (which can have a significantly different composition than the fragment) during close encounters within the Roche limit and could, therefore, acquire a very similar composition to the Earth.

In their study, \citet{kegerreisImmediateOriginMoon2022} focused on a narrow region of the parameter space close to the canonical model, and identified such immediately formed satellites for a substantive subset of simulations. In most cases, high resolution is required to reliably form these satellites. We observe such fragments in roughly \SI{3}{\percent} of the cases in our low resolution simulations over a wide range of the pre-impact parameter space ($\SI{1.0}{\JEM}\leq J_0\leq \SI{3.5}{\JEM}$, $0.04\leq\gamma\leq1.0$, $0.1\,v_{esc}\leq v_\infty \leq 1.0\, v_{esc}$ and $0.34\leq b_{\infty}\leq 1.09$) for all different rotation configurations and all angular velocity factors $f_\Omega$.

In Figure~\ref{fig:considerable_M_SLR_Disk_Mass}, the 191 cases with $M_b\geq\SI{1}{\Mearth}$ and a bound second largest remnant with a mass of $\SI{0.5}{\Mmoon} \leq M_{SLR} \leq \SI{1.5}{\Mmoon}$ are shown with the marker color being the mass of the SLR. Some of these fragments are embedded in relatively massive disks, while others comprise of nearly the full disk mass. In some cases, the disk finder does not count the fragment to the disk at all, because its orbit is such that it will eventually merge with the planet. These cases are excluded from Figure \ref{fig:considerable_M_SLR_Disk_Mass}.

Of the 191 results shown in Figure~\ref{fig:considerable_M_SLR_Disk_Mass}, 137 satisfy the permissive constraints ($M_b\geq\SI{1}{\Mearth}$, $M_d\geq\SI{1}{\Mmoon}$ and $F_d^{Fe}\leq0.04$) used in Figure~\ref{fig:successful_cases_loose}, while none satisfy the strict constraints. The disk iron mass fraction ranges from \SI{0.05}{\percent} (mostly rock) to \SI{27}{\percent} (iron rich) but there are no cases with zero disk iron mass fraction. The composition of the total disk (disk + fragment) ranges from the disk having a much lower impactor fraction than the planet (13 cases with $\delta_{pd} < 0.0$) to the disk having a much higher impactor fraction than the planet (178 cases with $\delta_{pd}>0.0$) with only 5 cases resulting in what we consider good mixing ($\vert\delta_{pd}\vert \leq 0.05$), all of which have $\gamma\geq 0.5$. All cases with low $J_b$ have a higher impactor fraction in the disk than in the planet. Disks that have a lower impactor fraction than the planet only occur for $J_b> \SI{2.0}{\JEM}$ and fragments with a much lower impactor fraction than the planet are only created in collisions with large impactor-to-target ratios ($\gamma \geq 0.9$) and either NX or DX configuration.

An interesting result is case 3567 which has pre-impact parameters close to the canonical model (NN, $J_0=\SI{1}{\JEM}$, $\gamma=0.1$, $v_{\infty}=0.2\, v_{esc}$, $b_{\infty}= 0.8153$) and produces a bound fragment of $M_{SLR}=\SI{0.84}{\Mmoon}$ with total disk mass of $M_d=\SI{0.99}{\Mmoon}$ at a bound angular momentum of $J_b=\SI{0.99}{\JEM}$, mixing parameter $\delta_{pd}=0.67$ and $F_d^{Fe}=0.0016$. A plot of the end state is shown in Figure~\ref{fig:NN_J1000_G0100_V02000_B08153_OMEGA0000}. While this case does not show adequate mixing between the proto-Earth and disk, the angular momentum of the system as well as the fragment mass and iron content are in excellent agreement with observational data. If the mixing constraint can be dropped (as suggested by \citealt{kaibBriefFollowupRecent2015,mastrobuono-battistiPrimordialOriginCompositional2015,mastrobuono-battistiCompositionSolarSystem2017,nielsenIsotopicEvidenceFormation2021}) this simulation would be an excellent match for the one that produced our Moon. 

If a massive bound fragment is embedded in the disk, it may act as a seed for Lunar formation and increase the fraction of the disk mass that will be accreted onto the Moon. In turn, if the disk is very massive and vapor dominated, the fragment could experience a drag force and spiral onto the proto-Earth (e.g., \citet{nakajimaLimitedRoleStreaming2024}). Clearly, the interplay between directly formed proto-satellites and a circumplanetary disk should be investigated in future work. Direct formation via giant impacts provides an interesting pathway for the formation of the Moon and other satellites. However, future work should investigate to which extent the numerical method could enhance direct formation and we suggest a follow-up study using higher resolution simulations (see also Appendix~\ref{sec:appendix_immediate}) to investigate, if these fragments can persist.

\subsection{Collision outcome degeneracy} \label{sec:result_degeneracy}

\begin{table}
\centering
\begin{tabular}{l|rr|rr|rr}
    \toprule
    Run & 3822 & 7334 & 2922 & 7605 & 2768 & 7563\\
    \midrule
    Orient. &NN&UU&DU&UU&DU&UU\\
    $J_0$ (\SI{}{\JEM}) &2.50&2.25&3.00&3.0&2.50&3.00\\
    $\gamma$ &0.5&0.3&1.0&0.7&0.5&0.3\\
    $v_{\infty}$ ($v_{esc}$) &1.0&0.4&0.8&0.9&0.1&0.8\\
    $b_{\infty}$ ($R_{grav}$) &0.542&0.570&0.685&0.337&0.962&0.416\\
    $f_{\Omega}$ &0.0&0.5&0.9&0.9&0.5&0.9\\
    \midrule
    $J_b$ (\SI{}{\JEM}) &2.240&2.235&2.980&2.950&2.492&2.580\\
    $M_d$ (\SI{}{\Mmoon})&2.098&2.174&4.438&4.451&3.421&3.490\\
    $F_d^{Fe}$ (\SI{}{\percent})&0.642&0.637&11.47&11.53&11.00&11.30\\
    $\delta_{pd}$ &0.559&0.564&0.107&0.108&0.404&0.425\\
    \bottomrule
\end{tabular}
\caption{Examples of collisions with different initial conditions that result in very similar outcomes discussed in Section~\ref{sec:result_degeneracy}. Such a degeneracy implies that one in principle cannot deduce the impact conditions from a specific collision outcome.}
\label{tab:Result_Degeneracy}
\end{table}

\begin{figure}[ht!]
\centering
\plotone{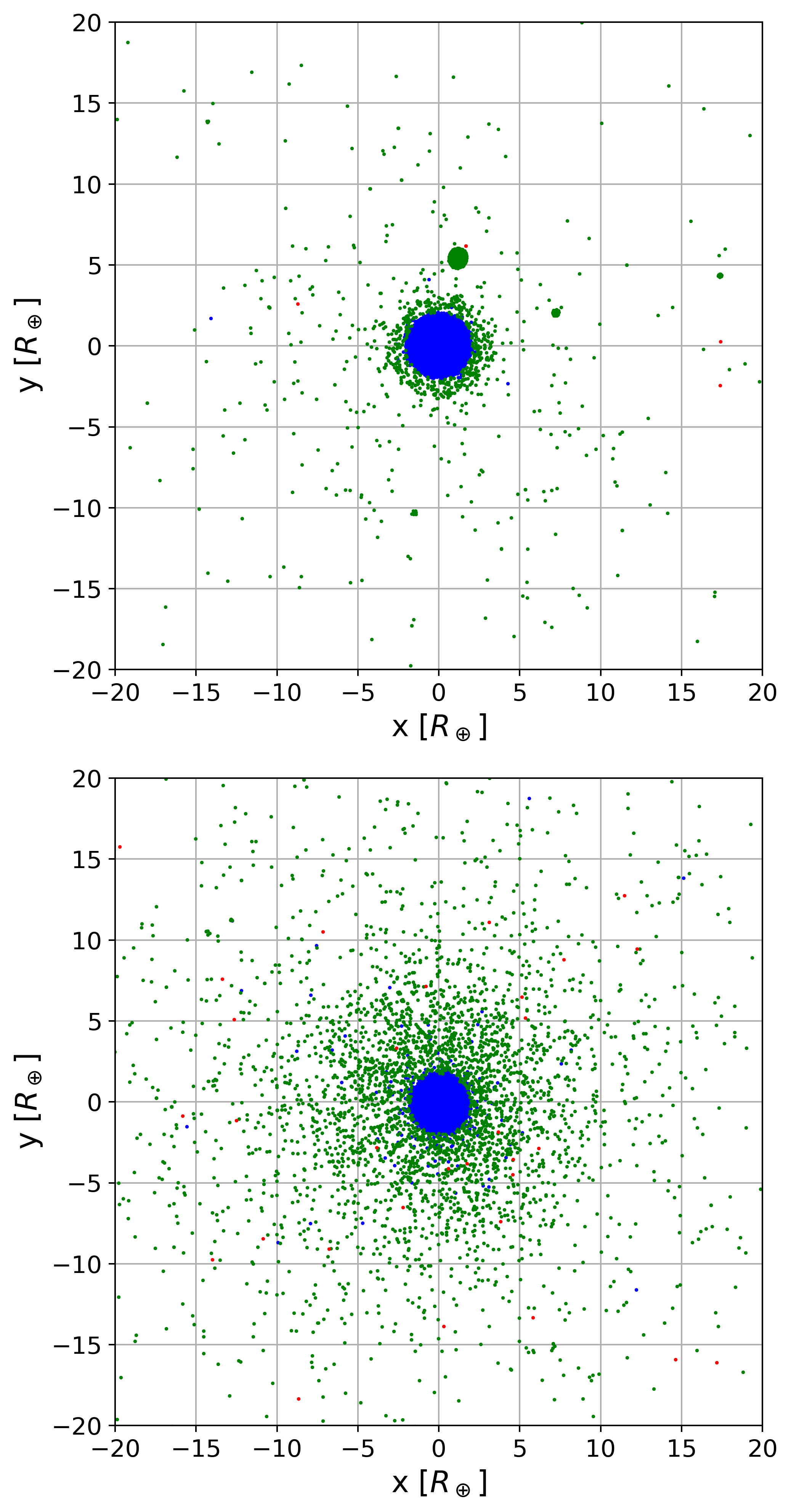}
\caption{Snapshots of two simulations with similar post-impact variables but significantly different disk morphology. Depicted is a projection of the particles into the plane of the collision. The particles are colored according to the result of the disk finder described in Appendix B in Paper I: blue (planet), green (disk) and red (ejected). The result of run 2768 (for pre-impact and post-impact parameters see Table \ref{tab:Result_Degeneracy}) in the top panel shows a massive bound second largest remnant of \SI{2.19}{\Mmoon}, while the run 7563 in the bottom panel does not have a bound fragment. They differ by only \SI{3.51}{\percent} in $J_b$, \SI{2.01}{\percent} in $M_d$, \SI{2.77}{\percent} in $F_{d}^{Fe}$ and \SI{5.04}{\percent} in $\delta_{pd}$. Thus, even for similar post-impact parameters the disk morphology can be very different.}
\label{fig:Similar_Results_With_and_Without_Bound_SLR}
\end{figure}

The simulation results show a large degeneracy with respect to the initial conditions, i.e., very different IC can result in almost identical collision outcomes. If we require the initial conditions for two collisions to be different in all parameters (orientation, angular momentum $J_0$, velocity $v_{\infty}$, impact parameter $b_{\infty}$ and angular velocity factor $f_\Omega$) we find for example the runs 3822 and 7334 (for pre-impact and post-impact parameters see Table \ref{tab:Result_Degeneracy}) that differ by only \SI{0.22}{\percent} in $J_b$, \SI{3.60}{\percent} in $M_d$, \SI{0.76}{\percent} in $F_{d}^{Fe}$ and \SI{0.82}{\percent} in $\delta_{pd}$. Allowing more similar IC, we find cases with even better agreement in the results. One example are the runs 2922 and 7605 where the $J_0$ and $f_{\Omega}$ are identical. The post-impact variables differ by only \SI{0.99}{\percent} in $J_b$, \SI{0.29}{\percent} in $M_d$, \SI{0.51}{\percent} in $F_{d}^{Fe}$ and \SI{1.11}{\percent} in $\delta_{pd}$. The outcome of these simulations can be considered identical within the typical accuracy of such simulations (e.g., \citealt{marcusCOLLISIONALSTRIPPINGDISRUPTION2009,canupLunarformingImpactsHighresolution2013,barrOriginEarthMoon2016,dengEnhancedMixingGiant2019}). 

Furthermore, even if the post-impact parameters of two collisions are very similar, the morphology of the resulting disks can be visibly different. As an example we show the result of two collision with very different IC in Figure~\ref{fig:Similar_Results_With_and_Without_Bound_SLR}. The post-impact variables (i.e., bound mass, disk mass and composition) differ by less than \SI{5.1}{\percent} in this case. However, the first (top panel) results in a massive bound second largest remnant (SLR) at the end of the simulation while the second (bottom panel) does not contain such a remnant.

We can conclude that the same result can in principle be generated with very different initial conditions. This means that even if one would find a collision which satisfies all constraints, it may still not be the impact that actually resulted in the formation of the Moon. In turn, one generally cannot determine (unique) pre-impact orbits of the proto-Earth and impactor, e.g., in order to constrain the pre-impact isotopic composition, from successful collisions.

\subsection{Stochasticity of impact outcomes}\label{sec:stochasticity}

\begin{figure*}[ht!]
\centering
\plotone{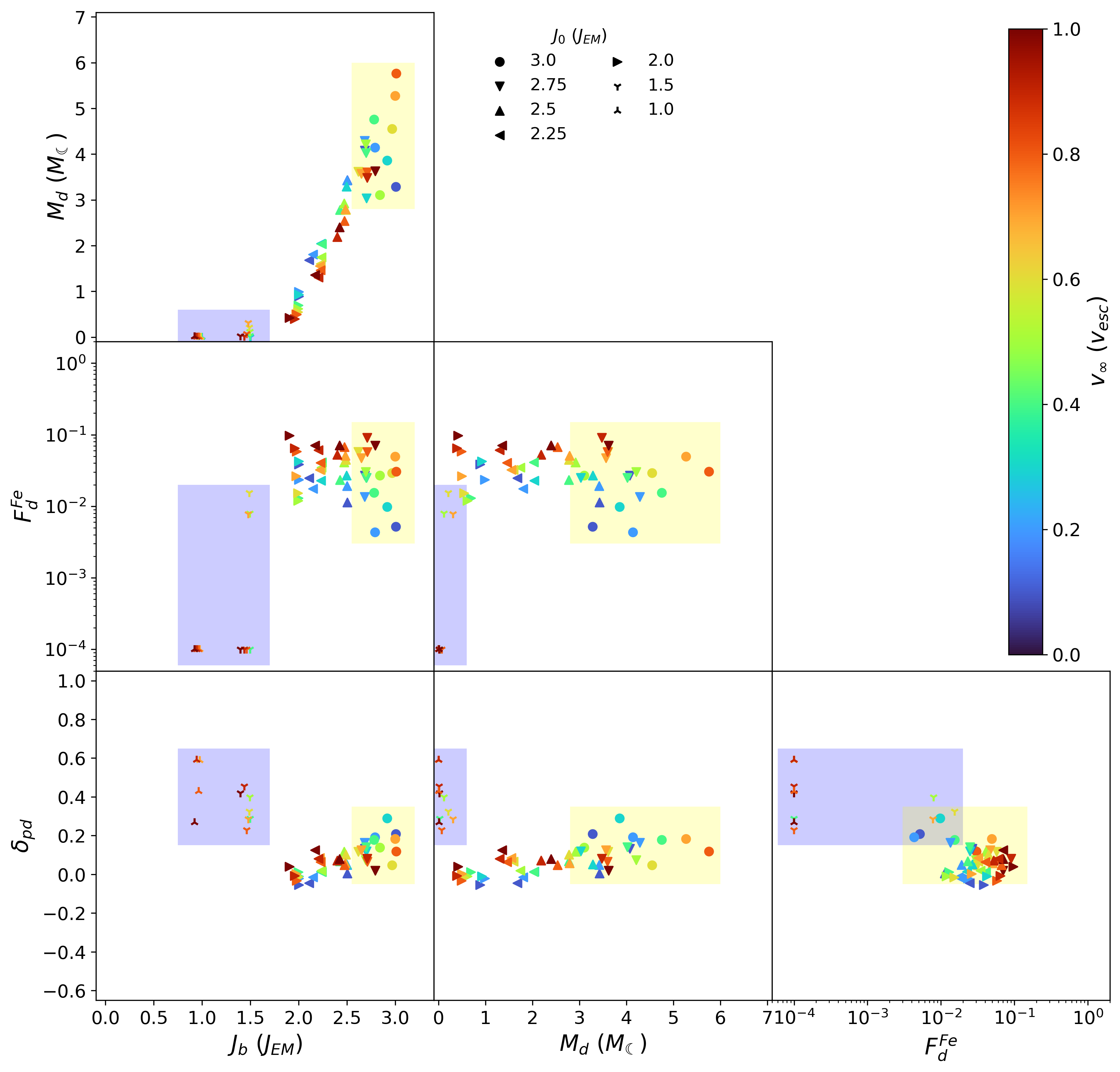}
\caption{An example set of collision results that illustrates the stochastic behavior in different regions of the parameter space.
Shown are 68 results for the NN configuration with $\SI{1.0}{\JEM}\leq J_0\leq\SI{3.0}{\JEM}$ and $\gamma=0.7$ constrained to $M_b\geq\SI{1}{\Mearth}$. Disk iron mass fractions of exactly zero are moved to \SI{1e-4}{} to be visible on a logarithmic plot (for details see caption of Figure~\ref{fig:Seaborn_Triangle_plot_only_output_variables}).
We observe three different regions: 1) The results in the region where $\SI{2.0}{\JEM}\leq J_0\leq\SI{2.5}{\JEM}$ exhibits smooth changes in the bound angular momentum $J_b$ and the disk mass $M_d$ and low scatter in the disk iron mass fraction $F_d^{Fe}$ and the mixing parameter $\delta_{pd}$. 2) For lower values of $J_0$ (blue shaded region), the disk mass is very low, resulting is large variations of both $F_d^{Fe}$ and $\delta_{pd}$. 3) For high values of $J_0$ (yellow shaded region), bound fragments occur leading to large variations in the post-impact values due to the stochastic nature of their formation (see Section~\ref{sec:stochasticity} for details).}
\label{fig:stochasticity}
\end{figure*}

The degeneracy in the results discussed above prompts the question, if the inverse, i.e., that very similar initial conditions can result in starkly different simulation outcomes, is also true. For the bulk of our simulations we do not find this to be the case. As an example, we show a small set of simulations in Figure \ref{fig:stochasticity}. If pre-impact parameters are changed slightly, the bound mass $M_b$, bound angular momentum $J_b$ and disk mass $M_d$ generally vary smoothly. The mixing parameter $\delta_{pd}$ and disk iron fraction $F_d^{Fe}$ are more stochastic, especially for low mass disks, that are poorly resolved and thus more prone to noise which makes them more sensitive to changes in initial conditions. Furthermore, at the boundaries between the different collision regimes, e.g., merger, hit-and-run or graze-and-merge, simulation outcomes are sensitive to small changes in the initial conditions (see also \citet{timpeMachineLearningApplied2020}) and possibly the resolution of the simulation. Finally, the presence of massive bound fragments introduces stochasticity to all disk parameters, including the disk mass, as the formation of fragments, their masses and orbital properties are sensitive to the choice of initial conditions and details of the numerical method. (see also Appendix \ref{sec:appendix_immediate} and \citet{kegerreisImmediateOriginMoon2022}). Future work should investigate stochasticity in GI simulations introduced by the formation of bound fragments and the question if it is reduced or completely vanishes at higher resolution.

\subsection{Summary}
Collisions between non-rotating bodies can only produce sufficiently massive disks ($M_d \geq \SI{}{\Mmoon}$) for post-impact angular momentum budgets of $J_b \gtrsim \SI{2}{\JEM}$, corresponding to an equivalent pre-impact angular momentum budget requirement ($J_0 \gtrsim \SI{2}{\JEM}$). Notably, small impactors ($\gamma < 0.1$) and $\gamma=0.1$ from the large impactor subset are unable to produce disks of at least one lunar mass, implying that the canonical Moon-forming impact is not viable in the absence of rotation. We also find that achieving compositional similarity between the proto-Earth and protolunar disk requires near-equal-mass mergers; near-perfect mixing can only be consistently achieved in equal-mass collisions.

In the present paper, we introduce pre-impact rotation to the colliding bodies. This increases the number of free parameters in our study by two (rotation configuration and angular velocity factor) and, therefore, requires significantly more simulations to fill the parameter space at the same resolution. Yet, despite this expansion of the pre-impact parameter space, we are still unable to identify a collision that is simultaneously consistent with the set of known constraints. Nonetheless, we obtained a number of useful insights into the parameter space that will inform future work on Moon-forming impacts. 

We find that even for pre-rotating bodies the disk mass and the post-impact angular momentum budget are strongly correlated. Generally, $J_b \gtrsim \SI{2}{\JEM}$ is required to produce disks of at least one lunar mass. However, if the proto-Earth is counter-rotating (DX configurations) with respect to the orbital angular momentum of the collision, disks with $M_d \geq \SI{1}{\Mmoon}$ and a bound angular momentum of $J_b \sim \SI{1}{\JEM}$ can be obtained. These cases only occur for collisions with $\gamma = 0.3 - 0.5$ and result in disks that have substantially higher impactor fraction than the planet. In order to explain the isotopic similarity between the proto-Earth and the Moon, such a scenario requires that either the impactor has a very similar isotopic composition or strong mixing between the proto-Earth and the disk occurred after the impact.

Good mixing requires near-equal mass ($\gamma \geq 0.5$) mergers or high velocity impacts with $\gamma = 0.025$ onto a rapidly spinning proto-Earth as proposed in \citet{canupFormingMoonEarthlike2012} and \citet{cukMakingMoonFastSpinning2012} respectively. Such collisions result in an excess angular momentum of \SIrange{1}{2}{\JEM}. Generally, if the disk must contain at least one lunar mass, one cannot reconcile the angular momentum of the Earth-Moon system with good mixing. Possible post-impact processes that could either result in mixing between the proto-Earth and the disk \citep{pahlevanEquilibrationAftermathLunarforming2007, lockOriginMoonTerrestrial2018} or efficiently remove the excess angular momentum \citep{cukMakingMoonFastSpinning2012, cukTidalEvolutionMoon2016, cukTidalEvolutionEarthMoon2021} were proposed but require very specific conditions. Furthermore, from a dynamical perspective the likelihood of equal mass mergers or high velocity collisions are unclear. N-body simulations investigating the Moon-forming impact within the planetesimal accretion paradigm suggest that such impact conditions are very rare \citep{kaibBriefFollowupRecent2015,kaibFeedingZonesTerrestrial2015}. However, simulations accounting for rapid oligarchic growth of planetary embryos closer to the Sun and including an early instability among the giant planets, suggest that equal mass mergers are more common \citep{clementEarlyInstabilityScenario2021}.

Significantly iron-depleted disks can be obtained over a very wide range of impact conditions and achieving a specific level of iron depletion is just a matter of fine-tuning. Thus, setting the constraint on the iron mass fraction aside, one can get two out of three: either massive disks with the correct AM or good mixing, or very low mass disks with the correct AM and good mixing.

We also find, that over a wide range of impact parameters relatively massive, bound fragments can form as a result of the collision. Such fragments result in about \SI{3}{\percent} of all simulations and can have very diverse properties. Among those 191 have masses between \SI{0.5}{} and \SI{1.5}{\Mmoon} and 131 satisfy the permissive constraints (see Section \ref{sec:Promising_Cases} for details). Furthermore, if a fragment is embedded in a disk, it could interact with the disk and act as a seed and enhance the accretion efficiency. While none of the simulations results in a (potential) satellite that matches the Moon, such a scenario could be an interesting pathway for lunar formation.

\section{Conclusions}
\label{sec:summary_and_conclusions}
We performed a systematic investigation of potential Moon-forming giant impacts. Our study consists of 7649 pairwise collisions between differentiated bodies with impactor-to-target mass ratios between $0.02$ and $1$ and nine distinct rotational configurations. This data set includes the 497 collisions between non-rotating bodies introduced in Paper I \citep{timpeSystematicSurveyMoonforming2023}.

\textbf{General observations:}
\begin{itemize}
    \item Despite the introduction of eight distinct rotation configurations and variable rotation rates, we can identify no single collision capable of simultaneously satisfying all known constraints. In all cases, one or more post-impact processes must be invoked to reconcile the constraints or it must be assumed that the target and impactor have the same isotopic composition prior to the impact.
    \item If the disk iron fraction constraint is ignored, we find that out of the remaining constraints ($M_d$, $J_b$ and $\delta_{pd}$), a maximum of two can be satisfied simultaneously, but never all three.
    \item Massive bound fragments (sometimes embedded in the disk) are a common outcome for a wide range of impact conditions and could be proto-satellites or act as seeds for accretion. 
\end{itemize}

\textbf{Systematic trends:}
\begin{itemize}
    \item The post-impact disk mass ($M_d$) remains strongly correlated to the post-impact angular momentum budget ($J_b$), with $J_b \gtrsim \SI{2}{\JEM}$ generally required to produce disks of at least one lunar mass. Thus, even with pre-impact rotation, the canonical Moon-forming impact is still incapable of producing favorably massive disks. However, a unique population of grazing, low-velocity impacts on counter-rotating targets at $0.3 \leq \gamma \leq 0.5$ breaks this relationship and produces low post-impact angular momentum budgets with massive disks.
    \item The disk iron mass fraction ($F^{Fe}_{d}$) is correlated with $v_{\infty}$ for large impactors ($\gamma \geq 0.1$), with higher impact velocities producing disks more enriched in iron. With pre-impact rotation, faster rotation rates are correlated with higher disk iron fractions. For small impactors, impacts on a co-rotating target (UN) produce significantly higher disk iron fractions than counter-rotating targets (DN).
    \item For large impactors that produce massive disks ($M_d \geq \SI{}{\Mmoon}$), compositional similarity between the post-impact planet and disk can only be achieved for high mass ratios ($\gamma \geq 0.5$) and becomes increasingly probable as $\gamma \to 1$. Notably, symmetry appears to play an important role in determining $\delta_{pd}$, with the symmetric rotation configurations (UU, NN, DD) consistently producing near-zero $\delta_{pd}$ for $\gamma=1$. For small impactors that produce massive disks, only collisions at $\gamma=0.025$ can satisfy the compositional constraint, however these cases result in excess angular momentum.
\end{itemize}

\textbf{Effect of rotation on leading theories:}
\begin{itemize}
    \item The canonical Moon-forming impact is only capable of producing lunar-mass disks ($M_d \geq \SI{}{\Mmoon}$) when the target is rapidly co-rotating (UX; $f_\Omega = 0.9$). This requires an excessive post-impact angular momentum budget ($J_b \gtrsim \SI{2}{\JEM}$) and the resulting disks evince substantial compositional differences with the proto-Earth ($\delta^{min}_{pd} > 0.2)$.
    \item Massive disks and very small compositional differences between the proto-Earth and the orbiting material can either be obtained by high velocity impacts of very small impactors onto rapidly co-rotating targets or near equal-mass mergers as proposed in \citet{cukMakingMoonFastSpinning2012} and \citet{canupFormingMoonEarthlike2012} respectively. However, those collisions result in an excess angular momentum of \SIrange{1}{2}{\JEM}.
    \item We identify a population of collisions that are uniquely capable of producing low post-impact angular momentum budgets and massive, iron-poor disks. This population represents a promising new class of Moon-forming impacts, but requires the target and impactor to have very similar compositions prior to the impact. In this scenario, a counter-rotating target roughly the mass of Venus suffers a grazing, low-velocity impact by an impactor roughly 2-3 times the mass of Mars.
\end{itemize}

While this study makes a first step towards understanding the systematics of Moon-forming impacts, there is still much work to be done. Future investigations should consider arbitrary mutual orientations, variable pre-impact core fractions, and further investigate the regions of interest identified in this study. Clearly, tighter constraints on the possible range of the post-impact state of the Earth-Moon system are required. Significant questions remain about accretion processes in the post-impact disk (e.g., accretion efficiency and possible enrichment in iron) and the efficacy of the proposed post-impact processes. To the latter point, simulations need to be done to constrain which post-impact states are suitable for various post-impact processes. Finally, connecting to formation models would allow to study lunar formation in a broader context and assess the frequency of satellites orbiting Earth-like planets.

\acknowledgments
We thank the anonymous reviewer for valuable suggestions and comments that helped to substantially improve the paper. We would also like to thank Martin Jutzi, Paolo Sossi, Maria Schönbächler and Jacob Kegerreis for helpful discussions. This work has been carried out within the framework of the National Centre of Competence in Research PlanetS supported by the Swiss National Science Foundation under grants 51NF40\_182901 and 51NF40\_205606. The authors acknowledge the financial support of the SNSF. We acknowledge access to Piz Daint and Eiger@Alps at the Swiss National Supercomputing Centre, Switzerland under the University of Zurich's share with the project ID UZH4.

%

\section*{Data Availability}
The data underlying this article is available in the Dryad Digital Repository: \href{https://doi.org/10.5061/dryad.8sf7m0czx}{doi:10.5061/dryad.8sf7m0czx}. The analysis results and an example Jupyter notebook to create figures from it can be found in the GitHub repository \citet{meierSurvey_Moon_Data2024} with the current version deposited to Zenodo: \href{https://zenodo.org/records/14060096}
{doi:10.5281/zenodo.14060096}.

\vspace{5mm}
\facilities{Swiss National Supercomputing Centre (Piz Daint, Eiger@Alps)}


\software{Gasoline \citep{wadsleyGasolineFlexibleParallel2004,reinhardtNumericalAspectsGiant2017},
          ballic \citep{reinhardtNumericalAspectsGiant2017},
          eoslib \citep{meierEOSlib2021,meierANEOSmaterial2021},
          skid \citep{n-bodyshopSKIDFindingGravitationally2011},
          numpy \citep{harrisArrayProgrammingNumPy2020},
          scipy \citep{virtanenSciPyFundamentalAlgorithms2020},
          matplotlib \citep{hunterMatplotlib2DGraphics2007},
          pynbody \citep{pontzenPynbodyNBodySPH2013},
          GNU parallel \citep{tangeGNUParallelCommandline2011}
          }



\appendix

\section{Specific pre-impact parameters}\label{sec:appendix_parameter_space}
As described in Section~\ref{sec:initial_conditions}, we restrict the orientation of the rotation axis of both the target and the impactor to three possible states: co-rotating with the orbital angular momentum (U), counter-rotating with the orbital angular momentum (D) and non-rotating (N). From these 3 states, we construct the rotation configuration of the collision (XX) where the first letter gives the orientation of the target and the second letter the orientation of the impactor. The possible rotation configurations are:

\begin{equation}
    \mathrm{XX} \in \left\{\mathrm{DD}, \mathrm{DN}, \mathrm{DU}, \mathrm{ND}, \mathrm{NN}, \mathrm{NU}, \mathrm{UD}, \mathrm{UN}, \mathrm{UU}\right\}.
\end{equation}

The parameter space presented in this paper is split into three regions:

\begin{itemize}
    \item 435 non-rotating (NN) collisions with large impactors ($\gamma \geq 0.1$) from Paper I,
    \item 4984 rotating collisions with large impactors ($\gamma \geq 0.1$),
    \item 2230 collisions with small impactors ($\gamma<0.1$),
\end{itemize}

\noindent where the first two regions together form the subset of large impactors (L) while the third region forms the subset of small impactors (S).

Given a rotation configuration, there are four free parameters that describe the initial conditions: the initial total angular momentum ($J_0$), the impactor-to-target mass ratio ($\gamma$), the asymptotic relative velocity ($v_{\infty}$) and the angular velocity factor ($f_{\Omega}$, if both bodies are rotating, they both have the same $f_\Omega$ value).

\subsection{Non-rotating collisions with large impactors}
The non-rotating (NN) simulations were the first simulations we ran and we used a high sampling resolution for $J_0$ to better understand the pre-impact parameter space. This sub-set contains all combinations of the following parameter choices that result in a collision (configurations that result in a fly-by were not run):

\begin{align}
    XX = &\,\mathrm{NN}\nonumber\\
    J_{0} \in &\left\{1.00, 1.50, 2.00, 2.25, 2.50, 2.75, 3.00, 3.25, 3.50, 4.00, 4.50, 5.00 \right\} \SI{}{\JEM}\nonumber\\
    \gamma \in &\left\{0.1, 0.3, 0.5, 0.7, 0.9, 1.0 \right\}\nonumber\\
    v_{\infty} \in &\left\{0.1, 0.2, 0.3, 0.4, 0.5, 0.6, 0.7, 0.8, 0.9, 1.0 \right\} v_{esc}\nonumber\\
    f_\Omega = &\,0.0
\end{align}

\noindent as well as additional simulations with $J_0=\SI{1.25}{\JEM}$ for $\gamma=0.1$, close to the canonical scenario. The results of these 435 simulations were thoroughly discussed in Paper I. 

\subsection{Rotating collisions with large impactors}
The first sub-set we add for Paper II consists of all combinations of these parameter choices that result in a collision:

\begin{align}
    XX \in &\left\{\mathrm{DD}, \mathrm{DN}, \mathrm{DU}, \mathrm{ND}, \mathrm{NU}, \mathrm{UD}, \mathrm{UN}, \mathrm{UU}\right\}\nonumber\\
    J_{0} \in &\left\{-1.75, -1.50, -1.25, -1.00, 1.00, 1.50, 2.00, 2.25, 2.50, 3.00 \right\} \SI{}{\JEM}\nonumber\\
    \gamma \in &\left\{0.1, 0.3, 0.5, 0.7, 0.9, 1.0 \right\}\nonumber\\
    v_{\infty} \in &\left\{0.1, 0.2, 0.3, 0.4, 0.5, 0.6, 0.7, 0.8, 0.9, 1.0 \right\} v_{esc}\nonumber\\
    f_\Omega \in &\left\{0.5, 0.9\right\}
\end{align}

\noindent This results in 4984 simulations. We use the same values for $\gamma$ and $v_{\infty}$ as for the non-rotating collisions. The values for $f_\Omega$ correspond to a body rotating at approximately \SI{50}{\percent} of the critical rotation rate, and a body rotating just shy of the critical rotation rate respectively. The body marked N in the NX and XN configurations is non-rotating. For the values of $J_0$, we decided to not exceed \SI{3.0}{\JEM}, even though we found in Paper I that mergers can be observed up to $J_0=\SI{3.5}{\JEM}$ for the NN configuration, because the bound AM is very similar to $J_0$ and removing such large amounts of AM is very difficult and thus these cases would not be relevant for the formation of the Moon. For the rotating configurations we also use a lower sampling resolution in $J_0$, because based on the NN results we are confident that the lower resolution is sufficient to identify systematic trends. The lower sampling resolution also prevents the study from becoming infeasible due to the additional computational resources required.

\begin{figure}[ht!]
\centering
\plotone{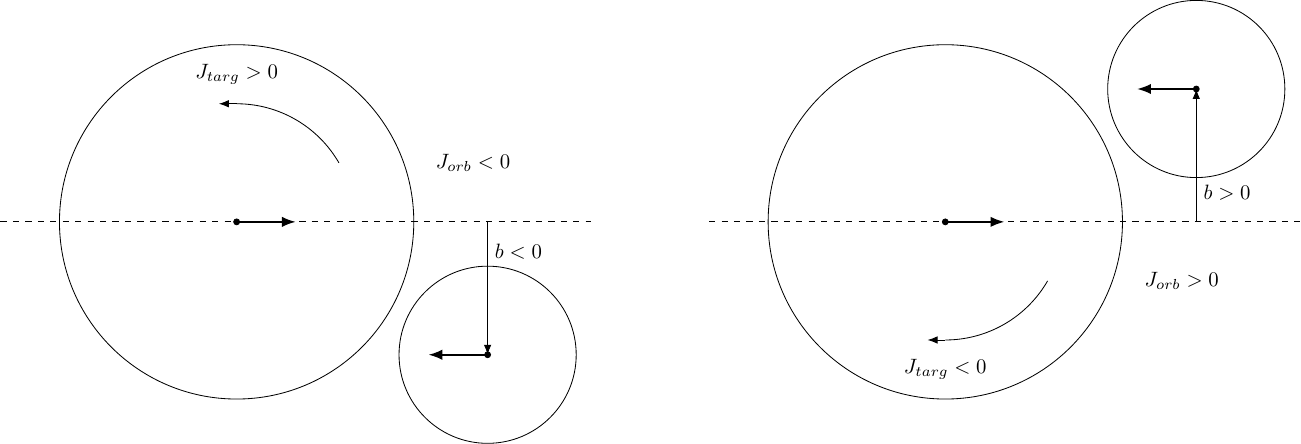}
\caption{An impact geometry with negative impact parameter $b$ can be transformed into a geometry with positive impact parameter by mirroring it on the plane perpendicular to $\vec{b}$. In this process, all angular momentum values change sign, such that the total initial angular momentum $J_0$ can be negative.}
\label{fig:Impact_Geometry}
\end{figure}

Our pipeline sets up the collisions such that the asymptotic impact parameter $b_{\infty}$ and, by extension, $J_{orb}$ is always positive. But, impacts with negative impact parameter are also part of the parameter space that could lead to the formation of the Moon. By mirroring the arrangement on the plane perpendicular to $\vec{b}$, such an impact with negative impact parameter can be transformed into one with positive impact parameter, while all angular momentum values change sign, such that the total initial angular momentum $J_0$ can be negative. This is shown in Figure~\ref{fig:Impact_Geometry} with the initial setup on the left side and the transformed setup with positive impact parameter on the right side.

Thus, in order to mimic collisions with negative impact parameter $b_{\infty}$, we add $J_0\in\{-1.75, -1.50, -1.25, -1.00\}\SI{}{\JEM}$ for the DX configuration (values below $J_0=\SI{-1.87}{\JEM}$ are not possible, because it is the smallest value of the sum $J_{rot,targ} + J_{rot,imp}$). In the NN case, the configuration is symmetric, but this is no longer the case if pre-impact rotation is added. It is not possible to create a negative $J_0$ value with the NX and UX configurations because $J_{orb}$ is always positive. The ND configuration naturally creates negative $J_0$, but the largest impactor ($\gamma=1.0$) rotating at $f_\Omega=0.9$ only has $J_{imp}=\SI{-0.71}{\JEM}$ which is not enough AM to get to a total angular momentum of \SI{-1.0}{\JEM}. The same is true for the UD case which could theoretically lead to a negative $J_0$, but this is not possible in our parametrization, as the absolute value of the angular momentum of a rotating target is always larger or equal to that of a rotating impactor.

\subsection{Collisions with small impactors}
In order to sample the parameter space of very fast counter-rotating targets, small impactors and high impact velocities proposed by \citet{cukMakingMoonFastSpinning2012}, we add a second sub-set for all orientations with non-rotating impactors (XN) which consists of all combinations of these parameter choices that result in a collision:

\begin{align}
    XX \in & \left\{\mathrm{DN}, \mathrm{NN}, \mathrm{UN}\right\}\nonumber\\
    J_{0} \in &\left\{-2.45, -2.25, -2.00, -1.75, -1.5, -1.25, -1.00,\right.\nonumber\\
    & \left. 1.00, 1.50, 2.00, 2.25, 2.50, 3.00, 3.50 \right\} \SI{}{\JEM}\nonumber\\
    \gamma \in &\left\{0.02, 0.025, 0.03, 0.04, 0.05\right\}\nonumber\\
    v_{\infty} \in &\left\{0.1, 0.2, 0.3, 0.4, 0.5, 0.6, 0.7, 0.8, 0.9, 1.0, \right.\nonumber\\
    & \left. 1.1, 1.2, 1.3, 1.4, 1.5, 1.6, 1.7, 1.8, 1.9, 2.0, \right.\nonumber\\
    & \left. 2.1, 2.2, 2.3, 2.4, 2.5, 2.6, 2.7, 2.8, 2.9, 3.0 \right\}v_{esc}\nonumber\\
    f_\Omega \in &\left\{0.0, 0.5, 0.9, 1.01\right\}
\end{align}

\noindent This results in 2230 simulations. For this subset, we increase the maximum asymptotic impact velocity $v_{\infty}$ from $1\,v_{esc}$ to $3\,v_{esc}$ and we add the angular velocity factor $f_{\Omega}=1.01$. This creates bodies rotating above $\Omega_{crit}$, which should not be stable, but because $\Omega_{crit}$ is an estimation assuming a homogeneous density, slightly larger values are still stable. $f_\Omega=1.01$ is the largest value that is stable for all our bodies. The value of $J_0 = \SI{-2.45}{\JEM}$ was chosen because $J_0 = \SI{-2.5}{\JEM}$ does not produce collisions for $\gamma\in\left\{0.04,0.05\right\}$. 

\subsection{Summary}
In total, we run 7649 impact simulations that are distributed as follows:

\begin{itemize}
    \item Rotational configuration: 2927 simulations with counter-rotating targets (DX), 1713 with non-rotating targets (NX) and 3009 with co-rotating targets (UX). A more detailed list of the number of simulations performed for each rotational configuration is provided in Table~\ref{tab:Simulation_Overview}.
    \item Initial total angular momentum $J_0$:
    \begin{eqnarray}
        1053 &\mbox{ with }& J_0=\SI{1.00}{\JEM}\quad 1000 \mbox{ with } J_0=\SI{1.50}{\JEM}\nonumber\\
        1080 &\mbox{ with }& J_0=\SI{2.00}{\JEM}\quad 1027 \mbox{ with } J_0=\SI{2.25}{\JEM}\nonumber\\
        1012 &\mbox{ with }& J_0=\SI{2.50}{\JEM}\quad 893 \mbox{ with } J_0=\SI{3.00}{\JEM}\nonumber\\
        111 &\mbox{ with }& J_0=\SI{3.50}{\JEM}\quad 1365 \mbox{ with negative } J_0\nonumber\\
        108 &\mbox{ with }& J_0\in\left\{1.25, 2.75, 3.25, 4.00, 4.50, 5.00\right\}\SI{}{\JEM}\nonumber
    \end{eqnarray}    
    \item Target-to-impactor mass ratio $\gamma$:
    \begin{eqnarray}
        431 &\mbox{ with }& \gamma=\SI{0.1}{}\quad 882 \mbox{ with } \gamma=\SI{0.3}{}\nonumber\\
        1009 &\mbox{ with }& \gamma=\SI{0.5}{}\quad 1034 \mbox{ with } \gamma=\SI{0.7}{}\nonumber\\
        1034 &\mbox{ with }& \gamma=\SI{0.9}{}\quad 1029 \mbox{ with } \gamma=\SI{1.0}{}\nonumber\\
        2230 &\mbox{ with }& \gamma<\SI{0.1}{} \mbox{ (low-}\gamma\mbox{ cases)}\nonumber
    \end{eqnarray}    
    \item Relative velocity at infinity $v_{\infty}$:
    \begin{eqnarray}
        526 &\mbox{ with }& v_\infty=0.1\,v_{esc}\quad 538 \mbox{ with } v_\infty=0.2\,v_{esc}\nonumber\\
        552 &\mbox{ with }& v_\infty=0.3\,v_{esc}\quad 569 \mbox{ with } v_\infty=0.4\,v_{esc}\nonumber\\
        590 &\mbox{ with }& v_\infty=0.5\,v_{esc}\quad 603 \mbox{ with } v_\infty=0.6\,v_{esc}\nonumber\\
        619 &\mbox{ with }& v_\infty=0.7\,v_{esc}\quad 628 \mbox{ with } v_\infty=0.8\,v_{esc}\nonumber\\
        642 &\mbox{ with }& v_\infty=0.9\,v_{esc}\quad 647 \mbox{ with } v_\infty=1.0\,v_{esc}\nonumber\\
        1735 &\mbox{ with }& v_\infty>1.0\,v_{esc}\nonumber
    \end{eqnarray}   
    \item Angular velocity factor $f_\Omega$: 497 non-rotating simulations ($f_\Omega=0.0$), 2755 simulations with $f_{\Omega} = 0.5$, 3494 simulations with $f_{\Omega} = 0.9$ and 903 simulations with $f_{\Omega} = 1.01$. We reiterate that the latter set is a specific subset of low-$\gamma$ collisions that was inspired by the parameter space proposed in \citet{cukMakingMoonFastSpinning2012}.
\end{itemize}

\section{Correlations in the full dataset} \label{sec:correlation_full_data_set}
\begin{figure}[ht!]
\centering
\plotone{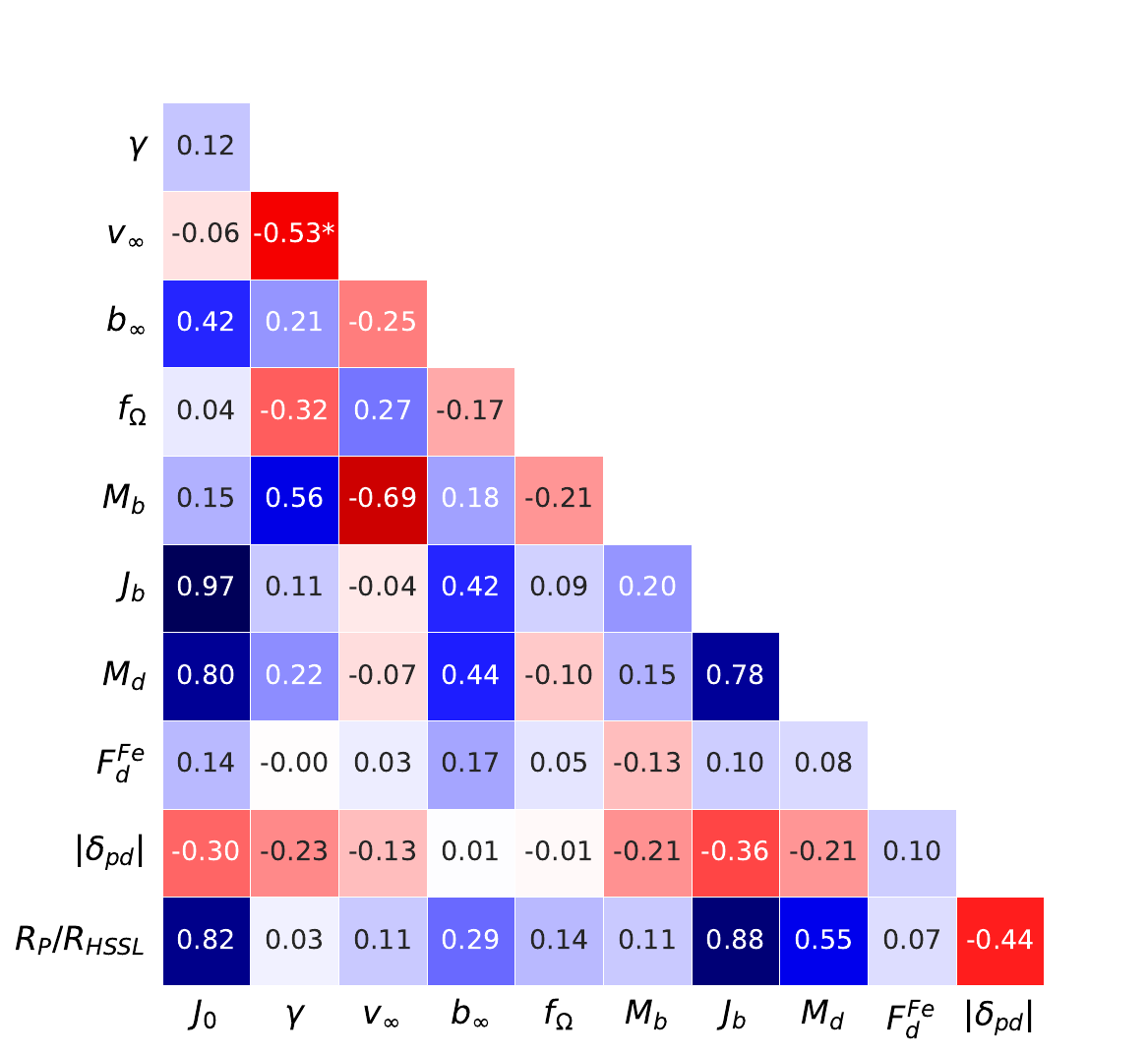}
\caption{Pearson correlation coefficients for a selection of pre-impact and post-impact properties of the 6247 simulations considered in this analysis. Blue squares indicate a positive correlation between properties, with stronger correlations marked by darker shades of blue. Red squares indicate a negative correlation, with darker shades of red indicating a stronger correlation. For a detailed discussion see Appendix~\ref{sec:correlation_full_data_set}.}
\label{fig:pearson_corr}
\end{figure}

In this appendix, we investigate global correlations between a subset of the pre-impact and post-impact properties. Figure~\ref{fig:pearson_corr} provides Pearson correlation coefficients for these properties. The coefficients are defined as

\begin{equation}
r_{XY}=\frac{\mathrm{cov}(X,Y)}{\sigma(X)\sigma(Y)} \,,
\end{equation}

\noindent where $\mathrm{cov}(X,Y)$ is the covariance and $\sigma(X)$ the standard deviation. The coefficients can range from $-1$ (perfect anti-correlation) to $1$ (perfect correlation). Values of $0.25 \leq \vert r_{XY} \vert \leq0.5$ are considered weak correlations while $\vert r_{XY} \vert>0.5$ are considered strong correlations. Note that the correlations between the pre-impact parameters can be affected by the way our initial conditions are generated and how the parameter space is sampled by our simulations (see Appendix \ref{sec:appendix_parameter_space}). 

We first look a the correlations between the pre-impact parameters ($J_0$, $\gamma$, $v_{\infty}$, $b_{\infty}$ and $f_\Omega$). They exhibit very low $r$ values, with the exception of $v_{\infty}$-$\gamma$ which evinces a coefficient of $r=-0.53$. This strong correlation is due to the fact that no collisions with $\gamma\geq0.1$ and $v_{inf}>1.0\,v_{esc}$ are simulated. $J_0$ and $v_{\infty}$ are not correlated because they are independent variables in our study. However, they are not perfectly uncorrelated because, for large values of $J_0$, high velocities can result in hit \& runs and low velocities are not simulated because they result in misses. The correlation between $J_0$ and $b_{\infty}$ is weak ($r=0.42$) because, for collisions with pre-impact rotation, $J_0$ contains both the orbital angular momentum and the spin angular momenta of the colliding bodies ($J_0 = J_{orb} + J_{targ} + J_{imp}$), which again are independent variables. Being an independent parameter, $f_\Omega$ should not show correlations with the other independent parameters, but nevertheless, it shows weak correlations with $\gamma$ ($r=-0.37$) and $v_{\infty}$ ($r=0.27$), because $f_\Omega=1.01$ is only used for $\gamma<0.1$ cases which mainly produce mergers for larger values of $v_{\infty}$.

Several strong correlations exist between pairs of pre-impact and post-impact parameters. The mass of the bound post-impact material ($M_b$) has a strong negative correlation of $r=-0.69$ with $v_{\infty}$. This is because high-velocity impacts tend to have small impact parameters (due to the initial angular momentum constraint) and the resulting head-on collisions tend to eject more material. $M_b$ is also strongly correlated with $\gamma$ ($r=0.56$), because collisions with high $\gamma$ are usually at lower impact velocities (see the negative correlation between $v_{\infty}$ and $\gamma$ mentioned above). The angular momentum of the bound mass $J_b$ correlates strongly ($r=0.97$) with $J_0$ because the majority of results (5896) exhibit a difference between $J_b$ and $J_0$ of less than \SI{20}{\percent} of $J_0$. This relation and its influence on the analysis is discussed in detail in Section~\ref{sec:Angular_Momentum}. $J_b$ also has a weak correlation with $b_{\infty}$ ($r=0.42$) because of the correlation between $b_{\infty}$ and $J_0$. $J_0$ also has a strong correlation ($r=0.80$) with the disk mass $M_d$, confirming a key finding of Paper I. From this correlation follows a strong correlation ($r=0.78$) between $J_b$ and $M_d$ which can be seen in the top-left panel of Figure~\ref{fig:all_other_paper_data_lines} and is consistent with the results of Paper I. $M_d$ also has a weak correlation ($r=0.44$) with $b_{\infty}$ because of its correlation with $J_0$.

The iron mass fraction of the disk $F_d^{Fe}$ does not show strong correlations with any parameter, the strongest value being $r=0.17$ with $b_{\infty}$. The absolute value of the mixing parameter $\vert\delta_{pd}\vert$ has no strong correlations with any of the other parameters, but has weak negative correlations with the pre-impact parameters $J_0$ and the post-impact parameter $J_b$ (because those are strongly correlated). Due to these correlations and the correlation between $J_b$ and $R_p/R_{HSSL}$, $\vert\delta_{pd}\vert$ also has a correlation with $R_p/R_{HSSL}$ of $r=-0.44$. It is interesting to note that, contrary to what one would expect if most of the impactor material is sheared into the disk, there is no correlation between $F_d^{Fe}$ and $\vert\delta_{pd}\vert$. This suggests that iron from the impactors core tends to fall back onto the proto-Earth in such collisions. 

The proximity of the post-impact planet to the hot-spin stability limit ($R_p/R_{HSSL}$) is strongly correlated with $J_0$ ($r=0.82$) and weakly correlated with $b_{\infty}$ ($r=0.29$). This is an intuitive result, as the radius of the planet ($R_p$) is determined by its rotation rate which increases with the amount of angular momentum in the post-impact system ($J_b$). Indeed, $J_b$ and $R_p/R_{HSSL}$ are strongly correlated ($r=0.88$) due to the strong correlation between $J_b$ and $J_0$. As for $b_{\infty}$, larger impact parameters are expected for large pre-impact angular momentum budgets, so it is unsurprising that $b_{\infty}$ and $R_p/R_{HSSL}$ share a weak correlation. 

\section{On the immediate formation of satellites}\label{sec:appendix_immediate}
As discussed in Section~\ref{sec:immediate_formation}, we find that many collisions result in a (more or less) massive second largest fragment which remains bound until the end of the simulation. In Figure~\ref{fig:considerable_M_SLR_Disk_Mass}, we show the 191 cases with fragment masses $\SI{0.5}{\Mmoon} \leq M_{SLR} \leq \SI{1.5}{\Mmoon}$ but our data set also contains 50 simulations with $M_{SLR} > \SI{1.5}{\Mmoon}$ with a maximum mass of the second largest fragment of \SI{3.2}{\Mmoon}. Such fragments are potential satellites or can act as a seed for the accretion of more mass from the circumplanetary disk and thus accelerate the formation and increasing the accretion efficiency. However, it is still being debated, to which extent the formation of such fragments can be enhanced by the numerical method. Known effects that can cause artificial fragmentation in SPH simulations are clumping due to the pairing instability \citep{dehnenImprovingConvergenceSmoothed2012}, regions with negative pressures in the EOS \citep{dehnenImprovingConvergenceSmoothed2012, zhangGeneralizedTransportvelocityFormulation2017} and the inherent graininess of the gravitational potential caused by discretization in particle based simulations \citep{bateResolutionRequirementsSmoothed1997}. \texttt{Gasoline}, the code used in the main part of this study should not suffer from the first two problems, as it uses the Wendland C2 kernel that is stable to pair instability, and it does not allow for negative pressures. \citet{kegerreisImmediateOriginMoon2022} (K22) shows that in certain cases, such immediately formed bound fragments could be physical, because, even though the final mass of the fragments varies slightly, the material flow leading to the formation of these fragments is stable to changes in resolution (above a certain minimum particle number needed to actually resolve the flow feature). They argue that it is possible that the Moon formed in such a scenario rather than being accreted from a disk. But they also caution that "the region of parameter space for the immediate formation of stable satellites is not huge". 

In this appendix we investigate this scenario and shed some light on the question if these stable fragments are physical or an artifact of their method specifically. For this, we try to reproduce the two most promising runs of \citet{kegerreisImmediateOriginMoon2022}, which they depict in the Figures 1 and 2 of their paper, using a newly developed SPH code based on pkdgrav3 (\citet{potterPKDGRAV3TrillionParticle2017}, Meier et al., in prep) at a resolution of $10^7$ particles. The SPH implementation in pkdgrav3 is based on \citet{springelCosmologicalSmoothedParticle2002} with corrections for $\nabla h$ terms, has an interface/surface correction based on \citet{ruiz-bonillaDealingDensityDiscontinuities2022} and uses ISPH \citep{reinhardtNumericalAspectsGiant2017} to enforce an adiabatic evolution in the absence of shocks. To avoid the pairing instability the high resolution Wendland C6 kernel with a target of 400 neighboring particles is used. Regarding the equations of state we follow K22 and use M-ANEOS $Fe_{85}Si_{15}$ \citep{stewartEquationStateModel2020} for the cores of the target and impactor and M-ANEOS Forsterite \citep{stewartEquationStateModel2019} for the mantles. To avoid artificial clumping we suppress negative pressures that can occur at low densities and temperatures.

\begin{figure*}[ht!]
\centering
\plotone{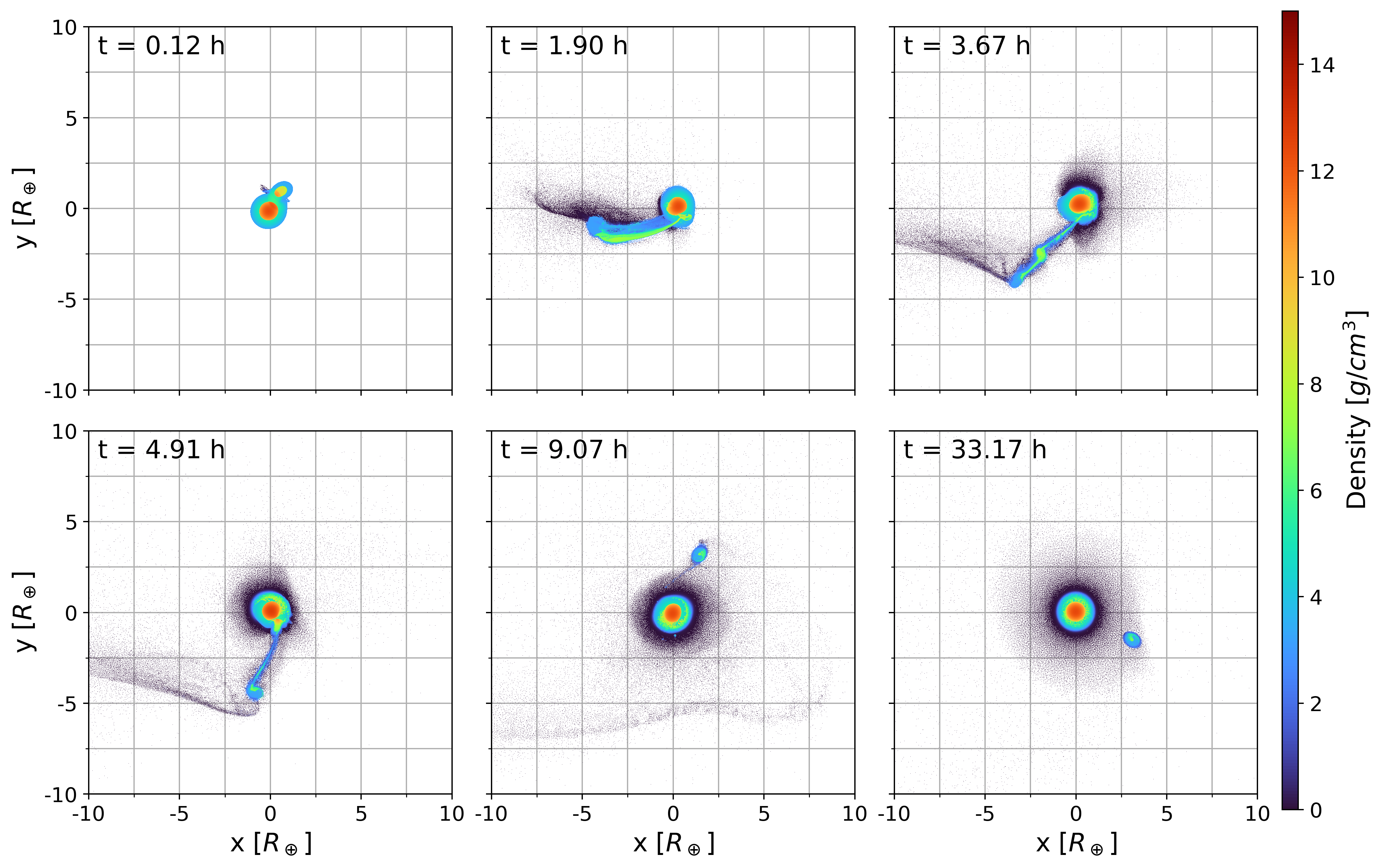}
\caption{Snapshots of our simulation with the initial conditions of the collision featured in Figure 2 of \citet{kegerreisImmediateOriginMoon2022}. Similar to them we form a massive satellite, but ours is more massive and on a lower orbit.}
\label{fig:Kegerreis_Successful}
\end{figure*}

\begin{figure*}[ht!]
\centering
\plotone{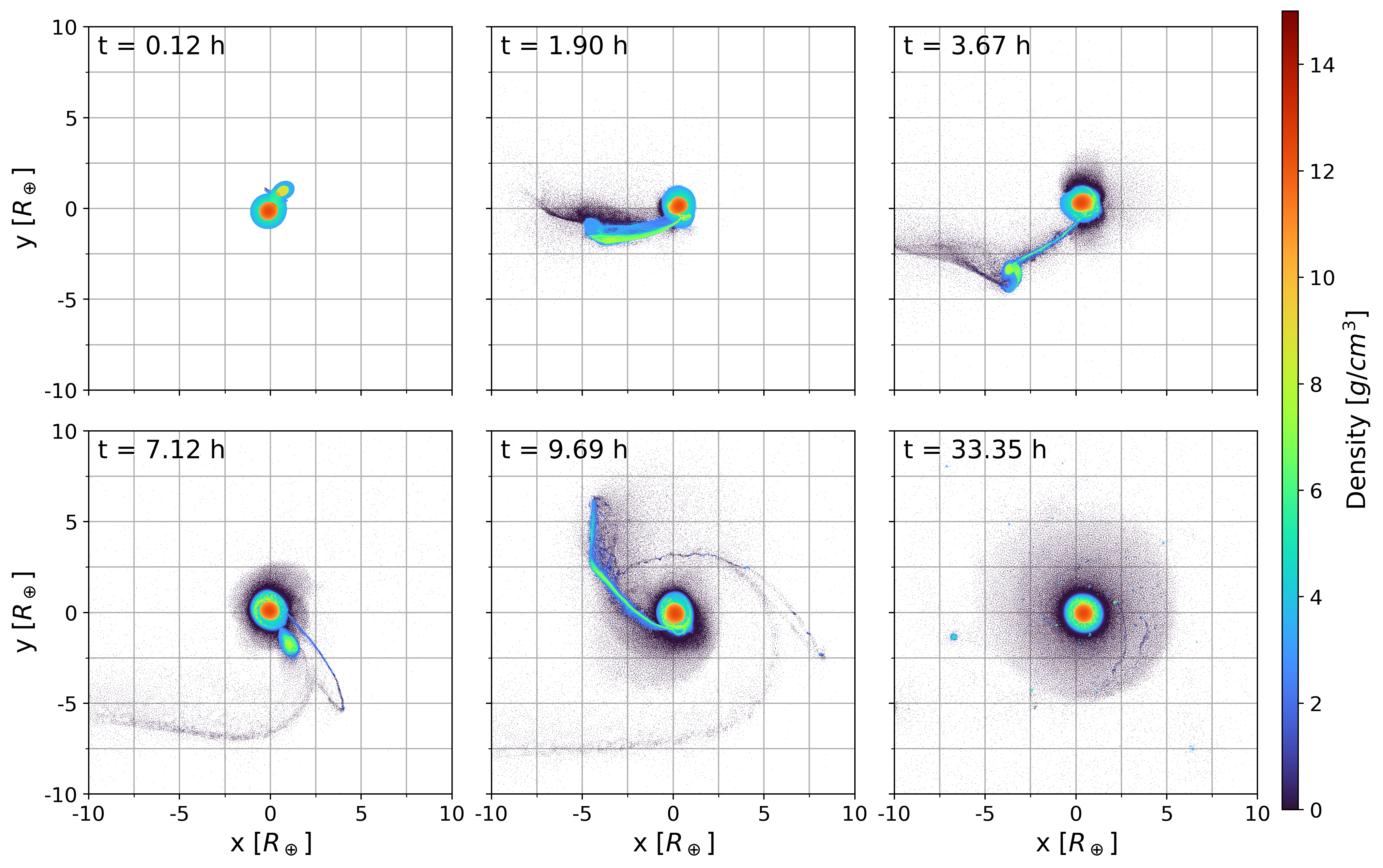}
\caption{Snapshots of our simulation with the initial conditions of the collision featured in Figure 1 of \citet{kegerreisImmediateOriginMoon2022}. Contrary to them, our simulation does not form a massive satellite, but shears the arm into a disk and many small satellites.}
\label{fig:Kegerreis_Unsuccessful}
\end{figure*}

In Figure~\ref{fig:Kegerreis_Successful} snapshots of a simulation with the initial conditions from Figure 2 of K22 are shown. Similar to their results, a massive satellite on a stable orbit forms in our simulation. However, in our simulations that satellite orbits closer to the Earth and has a mass of \SI{2}{\Mmoon} rather than \SI{1.41}{\Mmoon} as found in K22. Figure~\ref{fig:Kegerreis_Unsuccessful} shows snapshots of a simulation with the initial conditions of Figure 1 in K22. In this case, the fragment that forms from the arm in the top right panel falls back onto the proto-Earth, colliding in an oblique collision and gets sheared into another arm structure which then fragments into many very small bodies on varying orbits. These fragments are very low in mass with the most massive being of the order of a few percent of a Lunar mass.

In general, we can say that the concept of the immediate formation of satellites is possible. However, the exact properties, such as mass and initial orbit, of the resulting satellite seem very sensitive to small changes in initial conditions and details of the implementation of the numerical method. But this concept looks very promising and certainly warrants further investigation, especially because we find them all over the parameter space and at a (compared to \citealt{kegerreisImmediateOriginMoon2022}) small resolution (see Section~\ref{sec:immediate_formation}).





\bibliographystyle{aasjournal} 
\bibliography{main}






\end{document}